\documentclass[10pt,journal,romanappendices]{IEEEtran}
\usepackage{enumerate}
\usepackage{cite}
\usepackage{amsthm}
\usepackage{amsmath}
\usepackage{amscd}
\usepackage{dsfont}
\usepackage{bm}
\usepackage{amssymb}
\usepackage{latexsym}
\usepackage{array}
\usepackage{tikz}\usetikzlibrary{calc}
\usepackage{tabularx}
\usepackage[ruled]{algorithm}
\usepackage[noend]{algorithmic}
\usepackage{epsfig}
\usepackage{graphicx,subfigure}
\usepackage{epstopdf}
\usepackage[numbers, square, comma, compress]{natbib}
\newtheorem{lem}{Lemma}
\newtheorem{corol}{Corollary}
\newtheorem{ther}{Theorem}
\newtheorem{deft}{Definition}
\newtheorem{prop}{Proposition}

\theoremstyle{definition}
\newtheorem{rem}{Remark}
\newtheoremstyle{dotless}{}{}{\itshape}{}{\bfseries}{}{ }{}
\theoremstyle{dotless}

\newcommand{\I}{\mathbb{I}}
\newcommand{\Z}{\mathsf{Z}}
\newcommand{\M}{\mathsf{M}}
\newcommand{\V}{\mathbb{V}}
\DeclareMathOperator*{\Exp}{\mathbb{E}}
\newcommand{\R}{\mathbb{R}}

\newcommand{\X}{\mathsf{X}}
\newcommand{\Y}{\mathsf{Y}}
\newcommand{\D}{\mathbb{D}}
\newcommand{\h}{\mathbb{H}}
\newcommand {\aplt} {\ {\raise-.5ex\hbox{$\buildrel<\over{\mbox{\scriptsize $\sim$}}$}}\ }
\providecommand{\abs}[1]{\ensuremath{\left\lvert #1 \right\rvert}}
\providecommand{\norm}[1]{\ensuremath{\left\Vert #1 \right\Vert}}
\providecommand{\vv}[1]{\textquotedblleft #1\textquotedblright}
\providecommand{\nint}[1]{\ensuremath{\left\lfloor #1 \right \rceil}}
\providecommand{\floor}[1]{\ensuremath{\left\lfloor #1 \right \rfloor}}
\renewcommand{\IEEEQED}{\IEEEQEDopen}
\DeclareMathOperator*{\av}{av}

\DeclareMathOperator*{\Mod}{mod}
\newcommand{\Adv}{Adv}

\DeclareMathOperator*{\eff}{eff}
\DeclareMathOperator{\SNR}{\mathsf{SNR}}

\begin{document}

\title{Semantically Secure Lattice Codes \\for the Gaussian Wiretap Channel}

\author{Cong~Ling, Laura~Luzzi, Jean-Claude~Belfiore, and~Damien~Stehl\'{e}
\thanks{%
This work was supported in part by FP7 project PHYLAWS (EU FP7-ICT 317562), by a Royal Society-CNRS
international joint project and by a Marie Curie Fellowship (FP7/2007-2013, grant agreement PIEF-GA-2010-274765). This
work was presented in part at the IEEE International Symposium on
Information Theory (ISIT 2012), Cambridge, MA, USA.}
\thanks{%
C. Ling is with the Department of Electrical and
Electronic Engineering, Imperial College London, London SW7 2AZ,
United Kingdom (e-mail: cling@ieee.org).
\par
L. Luzzi was with the Department of Electrical and
Electronic Engineering, Imperial College London, London SW7 2AZ,
United Kingdom. She is now with
Laboratoire ETIS (ENSEA - Universit\'e de Cergy-Pontoise - CNRS), 6 Avenue du
Ponceau, 95014 Cergy-Pontoise, France (e-mail: laura.luzzi@ensea.fr).
\par
Jean-Claude Belfiore is with the Department of Communications and Electronics, Telecom ParisTech, Paris, France (e-mail: belfiore@telecom-paristech.fr).
\par
D. Stehl\'{e} is with ENS de Lyon, Laboratoire LIP (U.\ Lyon, CNRS, ENS de Lyon, INRIA, UCBL), 46 All\'ee
d'Italie, 69364 Lyon Cedex 07, France (e-mail: damien.stehle@ens-lyon.fr).}
}

\maketitle

\begin{abstract}
We propose a new scheme of wiretap lattice coding that achieves semantic security and strong secrecy over the Gaussian wiretap channel. The key tool in our security proof is the flatness factor which characterizes the convergence of the conditional output distributions
corresponding to different messages and leads to an upper bound on the information leakage.
We not only introduce the notion of secrecy-good lattices,
but also propose the {flatness factor} as a design criterion of such lattices. Both the modulo-lattice Gaussian channel and the genuine Gaussian channel are considered. In the latter case, we propose a novel secrecy coding scheme based on the discrete Gaussian distribution over a lattice, which achieves the secrecy capacity to within a half nat under mild conditions. No \textit{a priori} distribution of the message is assumed, and no dither is used in our proposed schemes.
\end{abstract}


\section{Introduction}

The idea of information-theoretic security stems from Shannon's
notion of \emph{perfect secrecy}. Perfect security can be achieved
by encoding an information message~$\M$ (also called plaintext message),
belonging to a finite space~$\mathcal{M}$,
into a codeword or ciphertext~$\Z$, belonging to a discrete or continuous space~$\mathcal{Z}$,
in such a way that the mutual information $\I(\M;\Z)=0$. However, perfect
security is not so practical because it requires a one-time pad.

In the context of noisy channels, Wyner \cite{Wyner75} proved that
both robustness to transmission errors and a prescribed degree of
data confidentiality could simultaneously be attained by channel
coding without any secret key. Wyner replaced Shannon's perfect
secrecy with the \emph{weak secrecy} condition $\lim_{n\rightarrow
\infty} \frac{1}{n}\I(\M;\Z^n)= 0$, namely the asymptotic rate of
leaked information between the message $\M$ and the channel
output~$\Z^n$ should vanish as the block length $n$ tends to infinity.

Unfortunately, it is still possible for a scheme satisfying weak
secrecy to exhibit some security flaws, e.g., the total amount of
leaked information may go to infinity, and now it is widely accepted
that a physical-layer security scheme should be secure in the sense
of Csisz\'ar's \emph{strong secrecy} $\lim_{n\rightarrow \infty}
\I(\M;\Z^n)= 0$ \cite{Csiszar96}.

In the notion of strong secrecy, plaintext messages are often assumed to be
random and uniformly distributed in~$\mathcal{M}$. This assumption is deemed
problematic from the cryptographic perspective, since in many setups plaintext
messages are not random. This issue can be resolved by using the
standard notion of \emph{semantic security}~\cite{GoMi84} which
requires that the probability that the eavesdropper can guess any
function of the message given the ciphertext should not be
significantly higher than the probability of guessing it using a simulator that
does not have access to the ciphertext.
The relation between
strong secrecy and semantic security was recently revealed in~\cite{Bellare_Tessaro_Vardy}
for discrete wiretap channels,
namely, achieving strong secrecy
for all distributions of the plaintext messages is equivalent to achieving
semantic security.

Wiretap codes achieving strong secrecy over discrete memoryless channels have been proposed in
\cite{Suresh_2010,Mahdavifar_Vardy_2011}. In particular, polar codes
in \cite{Mahdavifar_Vardy_2011} also achieve semantic security (this was implicit in \cite{Mahdavifar_Vardy_2011}), although reliability over the main channel is not proven when it is noisy. For
continuous channels such as the Gaussian channel, the problem of
achieving strong secrecy has been little explored so far and the design of wiretap codes
has mostly focused on the maximization of the eavesdropper's error probability \cite{Klinc2011}.
Recently, some progress
has been made in wiretap lattice codes over Gaussian wiretap
channels. It is quite natural to replace Wyner's random binning with
coset coding induced by a lattice partition $\Lambda_e \subset
\Lambda_b$. The secret bits are used to select one coset of the
coarse lattice $\Lambda_e$ and a random point inside this coset is
transmitted.
Wiretap lattice codes from an error probability point of view
were proposed in \cite{OSB11}, which also introduced the notion of \emph{secrecy
gain} and showed that the eavesdropper's error probability
$\lim_{n\rightarrow \infty} P_e=1$ for even unimodular lattices. These lattice codes
were further investigated in \cite{Ernvall-Hytonen_Hollanti2011}.
In \cite{CLW11} the existence of wiretap lattice codes (based on the
ensemble of random lattice codes) achieving the secrecy capacity
under the weak secrecy criterion was demonstrated.
Finally, we note that the secrecy capacity of the continuous mod-lattice channel
with feedback was studied in \cite{Lai08}, and that standard lattices codes
for the Gaussian channel \cite{ErezZamir04} were used to provide weak/strong secrecy
in the settings of cooperative jamming and interference channels in \cite{HeYener08,HeYener11,HeYener13}.

\subsection*{Main Contributions}

In the present work, we propose wiretap lattice codes that achieve strong
secrecy and semantic security over (continuous) Gaussian wiretap channels.
Firstly, we extend the relation between strong
secrecy and semantic security \cite{Bellare_Tessaro_Vardy}
to continuous wiretap channels. We further derive a bound on
the mutual information in terms of the variational distance for
continuous channels. More importantly, we propose the \emph{flatness factor} of a lattice
as a fundamental criterion which guarantees $L^1$
convergence of conditional outputs and characterizes the amount of information leakage.
This leads to the definition of \vv{secrecy-good lattices}. Letting the coarse lattice $\Lambda_e$ be secrecy-good, we straightforwardly tackle the problem of secrecy coding for the \emph{mod-$\Lambda$
wiretap channel}.
We then extend the scheme to the Gaussian wiretap scenario by employing \emph{lattice Gaussian
coding}. More precisely, the
distribution of each bin in our wiretap code is a discrete Gaussian
distribution over a coset of a secrecy-good lattice. We use
the flatness factor to show that this scheme can approach the secrecy capacity of the Gaussian
wiretap channel up to a constant gap of $\frac{1}{2}$ nat (under very mild assumptions),
by using minimum mean-square error (MMSE) lattice decoding at
the legitimate receiver.

The proposed approach enjoys a couple of salient features. Firstly, throughout the paper, we do not make any assumption on
the distribution of the plaintext message~$\M$, i.e., the security holds for any particular message. Thus, similarly to
\cite{Bellare_Tessaro_Vardy}, the proposed wiretap lattice codes can
achieve semantic security. Secondly, in contrast to \cite{ErezZamir04}, we do not use a dither. This may simplify the implementation of the system.

\subsection*{Relations to Existing Works}

\emph{Relation to secrecy gain:} The secrecy gain ~\cite{OSB11} is based on the error probability analysis, while the flatness factor is directly related to the variational distance and mutual information. Yet, despite the different criteria, they are both determined by the theta series and are in fact consistent with each other. Given the fundamental volume of a lattice, a small flatness factor requires a
small theta series, which coincides with the criterion from~\cite{OSB11}
for enjoying a large secrecy gain.

\emph{Relation to resolvability:} In \cite{Bloch11, LuzziBloch11}, a
technique based on \emph{resolvability} was suggested to obtain strong
secrecy, which uses a binning scheme such that the bin rate is above
the capacity of the eavesdropper's channel. We will show this is
also the case for the proposed lattice scheme.

\emph{Relation to lattice-based cryptography:}
Lattice-based cryptography~\cite{MicciancioRegev08} aims at realizing classical cryptographic
primitives, such as digital signatures and public-key encryption schemes, that
are provably secure \emph{under algorithmic hardness assumptions} on worst-case
lattice problems, such as variants of the decisional shortest vector problem.
In the present work, we propose an encryption scheme without keys for the Gaussian wiretap
channel that involves lattices, but the security is proven \emph{without algorithmic hardness assumptions}.\\

\subsection*{Organization}

Section II studies the relation between semantic security and strong secrecy for continuous wiretap channels. In Section III, we review lattice Gaussian distributions and propose the flatness factor as a novel machinery.
Sections IV and V address the mod-$\Lambda$ channel and the Gaussian wiretap channel, respectively.
In Section VI, we conclude the paper with a brief discussion of open issues.

Throughout this paper, we use the natural logarithm, denoted by $\log$, and information is measured in nats. We use the standard asymptotic notation $%
f\left( x\right) =O\left( g\left( x\right) \right) $ when $\lim\sup_{x\rightarrow
\infty}|f(x)/g(x)| < \infty$, $f\left( x\right) =\Omega \left( g\left( x\right) \right) $ when
$\lim\sup_{x\rightarrow \infty}|g(x)/f(x)| < \infty$, $%
f\left( x\right) =o\left( g\left( x\right) \right) $ when $\lim\sup_{x\rightarrow
\infty}|f(x)/g(x)| =0$, and $%
f\left( x\right) =\omega\left( g\left( x\right) \right) $ when $\lim\sup_{x\rightarrow
\infty}|g(x)/f(x)| =0$ .

\section{Strong secrecy and semantic security in continuous channels}

In this section, we investigate the relation between strong secrecy
and semantic security in continuous  wiretap channels.

\subsection{Wiretap Codes}
Consider an $n$-dimensional continuous memoryless wiretap channel with input $\X^n$
and outputs $\Y^n$, $\Z^n$
for the legitimate receiver and  the eavesdropper respectively.

\begin{deft}[Wiretap code \cite{Bloch_Barros_2011,Liang_Poor_Shamai_2009}]
An $(R,R',n)$ wiretap code is given by
a message set $\mathcal{M}_n=\{1,\ldots,e^{nR}\}$, an auxiliary
discrete random source $\mathsf{S}$ of entropy rate $R'$ taking values
in $\mathcal{S}_n$,
an encoding
function $f_n: \mathcal{M}_n \times \mathcal{S}_n \to \R^n$ and a
decoding function $g_n: \R^n \to \mathcal{M}_n$ for the legitimate
receiver. Let $\X^n=f_n(\M,\mathsf{S})$ be the channel input for a distribution~$\M$
of messages, and
$\hat{\M}=g_n(\Y^n)$ the estimate of the legitimate receiver.

\end{deft}

There are two options to define the transmission power:
\begin{itemize}

  \item Average power constraint: Channel input $\X^n$ satisfy the
constraint
\begin{equation} \label{power_constraint}
\frac{1}{n}\mathbb{E} \left[ \norm{\X^n}^2\right] \leq P
\end{equation}
with respect to~$\M$ chosen as the uniform distribution and to the randomness source~$\mathsf{S}$.

  \item Individual power constraint: One can impose a more stringent power constraint on each individual bin (without assuming $\M$ is uniformly distributed):
\begin{equation} \label{bin_power_constraint}
\forall m \in \mathcal{M}_n, \quad \frac{1}{n}\mathbb{E}_{\mathsf{S}}  \left[\norm{f_n(m,\mathsf{S})}^2\right] \leq P.
\end{equation}
\end{itemize}

Incidentally, the proposed lattice codes will satisfy the individual power constraint.

\subsection{Strong Secrecy and Semantic Security}

The {Kullback-Leibler divergence} of the distributions
$p_{\X}$ and~$p_{\Y}$ is defined as~$\D(p_{\X} \Vert p_{\Y})=\int_{\R^n}
p_{\X}(\mathbf{x}) \log
\frac{p_{\X}(\mathbf{x})}{p_{\Y}(\mathbf{x})} d\mathbf{x}$. The mutual information between
$\X,\Y$ is defined by
\begin{equation*}
\I(\X;\Y)=\D(p_{\X\Y} \Vert p_{\X}p_{\Y}).
\end{equation*}
The  \emph{variational distance} or
\emph{statistical distance} is defined by
\begin{equation*}
\V(p_{\X},p_{\Y}) \triangleq \int_{\R^n} \abs{p_{\X}(\mathbf{x})-p_{\Y}(\mathbf{x})}d\mathbf{x}.
\end{equation*}

With the definitions given above, we are ready to introduce strong
secrecy and semantic security.

\begin{deft}[Achievable strong secrecy rate]
The message rate $R$ is an \emph{achievable strong secrecy rate} if
there exists a sequence of wiretap codes $\{\mathcal{C}_n\}$ of rate
$R$ such that
\begin{align}
&\mathbb{P}\{\hat{\M} \neq \M\}  \to 0, \tag*{(reliability)}\\
&\mathbb{I}(\M;\Z^n) \to 0 \tag*{(strong secrecy)}
\end{align}
when $n \to \infty$.
\end{deft}

In the definition of strong secrecy for communications, no special attention is paid to the issue of message distribution.
In fact, a uniform distribution is often assumed in the coding literature.
But this is insufficient from a cryptographic viewpoint, as it does not ensure
security for a particular message. To address this issue of the wiretap code, we need to ensure the mutual information vanishes for all message distributions:
\begin{equation} \label{mis}
\mathrm{Adv^{mis}}(\Z^n) \triangleq \max_{p_{\M}} \mathbb{I}(\M;\Z^n) \to 0
\end{equation}
when $n \to \infty$. The \emph{adversarial advantage} $\mathrm{Adv^{mis}}$ tending to zero was termed \emph{mutual information security} in \cite{Bellare_Tessaro_Vardy}. In this paper, the terms mutual information security and strong secrecy for all message distributions are used interchangeably. Note that one may further impose constraints on the rate of convergence towards~$0$; in practice an exponential rate of convergence is desired.

Let the min-entropy of a discrete random variable $\M$ be
\[\mathbb{H}_{\infty}(\M)=-\log\left(\max_m \mathbb{P}\{\M=m\}\right).\]

\begin{deft}[Semantic security]
A sequence of wiretap codes~$\{\mathcal{C}_n\}$ achieves semantic
security if
\begin{align*}
&\mathrm{\Adv^{ss}}(\Z^n) \triangleq \sup_{f,{p_{\M}}} \left( e^{-\mathbb{H}_{\infty}(f(\M)|\Z^n)} -
e^{-\mathbb{H}_{\infty}(f(\M))} \right) \to 0
\end{align*}
when $n \to \infty$. The supremum is taken over all message distributions~$p_{\M}$
 and all functions $f$ of $\M$ taking values in the set $\{0,1\}^*$ of finite
binary words.
\end{deft}

Semantic security means that, asymptotically, it is impossible to
estimate any function of
the message better than to guess it without considering~$\Z^n$ at all.
We also define distinguishing security, which means that,
asymptotically, the channel outputs are indistinguishable for
different input messages.

\begin{deft}[Distinguishing security]
A sequence of wiretap codes $\{\mathcal{C}_n\}$ achieves
distinguishing security if
\begin{align}\label{DS}
&\mathrm{Adv^{ds}}(\Z^n) \triangleq \max_{m,m'}\mathbb{V}(p_{\Z^n|\M=m},p_{\Z^n|\M=m'}) \to 0,
\end{align}
when $n \to \infty$.
The maximum in the previous equation is taken over all messages $m,m' \in \mathcal{M}_n$.
\end{deft}

As for the discrete wiretap channel setup considered in~\cite{Bellare_Tessaro_Vardy},
the classical proof of equivalence between semantic security and distinguishing
security~\cite{GoMi84} can be readily adapted and it can be shown that\footnote{Note that
the factors in~\cite{Bellare_Tessaro_Vardy} are 1 on the left and 2 on the right, respectively,
due to the factor $\frac{1}{2}$ used in the definition of the variational distance in~\cite{Bellare_Tessaro_Vardy}.}
\begin{equation} \label{ss_ds_equivalence}
2\mathrm{Adv^{ss}}(\Z^n) \leq \mathrm{Adv^{ds}}(\Z^n) \leq 4\mathrm{Adv^{ss}}(\Z^n).
\end{equation}
Even though the two definitions are equivalent, distinguishing security often turns out to be technically easier to manipulate.

\subsection{Equivalence}

We will show that semantic security and strong secrecy for all message distributions are equivalent for
continuous channels. This is an extension of the results
from Section~3 of~\cite{Bellare_Tessaro_Vardy}.

We first need the following continuous channel adaptation of
Csisz\'ar's in~\cite[Lemma~1]{Csiszar96}. The lower bound is a consequence of
Pinsker's inequality (see~\cite[pp.58-59]{CsiszarKorner81}).
The proof of the upper bound is similar to the discrete case and is given in Appendix~\ref{Csis}.

\begin{lem} \label{Csiszar_lemma}
Let $\Z^n$ be a random variable defined on $\R^n$
and~$\M$ be a random variable over a finite domain~$\mathcal{M}_n$
such that~$|\mathcal{M}_n| \geq 4$.
Then
\[
\frac{1}{2}  d_{\av}^2 \leq \I(\M;\Z^n) \leq d_{\av} \log \frac{|\mathcal{M}_n|}{d_{\av}},
\]
where
\[
d_{\av} =\sum_{m \in \mathcal{M}_n} p_{\M}(m) \V(p_{\Z^n|\M=m},p_{\Z^n})
\]
is the average variational distance of the conditional output distributions from the global output distribution.
\end{lem}

We now prove the equivalence between semantic security and strong secrecy for all message distributions via distinguishing security.
\begin{prop}
a) A sequence of wiretap codes $\{\mathcal{C}_n\}$ of rate $R$ which achieves semantic security with advantage $\mathrm{Adv^{ds}}(\Z^n)=o\left(\frac{1}{n}\right)$  also achieves strong secrecy for all message distributions, namely, for all $p_{\M}$,
\[
\I(\M;\Z^n) \leq \mathrm{Adv^{mis}}(\Z^n) \leq \varepsilon_n  \left(nR-\log \varepsilon_n\right),
\]
where~$\varepsilon_n \triangleq \mathrm{Adv^{ds}}(\Z^n)$.
b) A sequence of wiretap codes $\{\mathcal{C}_n\}$ which achieves strong secrecy for all message distributions also achieves semantic security:
\[
\mathrm{Adv^{ds}}(\Z^n) \leq 2\sqrt{2 \mathrm{Adv^{mis}}(\Z^n)}.
\]
\end{prop}

\begin{IEEEproof}


(a) Distinguishing security $\Rightarrow$ strong secrecy for all message distributions: For any~$m \in \mathcal{M}_n$, we have
{
\allowdisplaybreaks
\begin{align*}
&\V(p_{\Z^n|\M=m},p_{\Z^n}) \\
&= \int_{\R^n} \Big| p_{\Z^n|\M}(\mathbf{z}|m)-\sum_{m' \in \mathcal{M}_n} p_{\M}(m')p_{\Z^n|\M}(\mathbf{z}|m')\Big| d\mathbf{z} \\
&=\int_{\R^n} \Big|\sum_{m' \in \mathcal{M}_n} p_{\M}(m') \left(p_{\Z^n|\M}(\mathbf{z}|m)- p_{\Z^n|\M}(\mathbf{z}|m')\right)\Big| d\mathbf{z} \\
& \leq \max_{m' \in \mathcal{M}_n}\mathbb{V}(p_{\Z^n|\M=m},p_{\Z^n|\M=m'}) \\
& \leq \max_{m',m'' \in \mathcal{M}_n}\mathbb{V}(p_{\Z^n|\M=m'},p_{\Z^n|\M=m''})=\varepsilon_n.
\end{align*}}
Therefore $d_{\av} \leq \varepsilon_n$. By 	
Lemma~\ref{Csiszar_lemma}, we obtain
\begin{equation*}
\I(\M;\Z^n) \leq \varepsilon_n \log \frac{|\mathcal{M}_n|}{\varepsilon_n} =
\varepsilon_n nR -\varepsilon_n\log\varepsilon_n.
\end{equation*}
If $\mathrm{Adv^{ds}}(\Z^n)=o(\frac{1}{n})$, then $\I(\M;\Z^n)\rightarrow 0$.

\smallskip

(b) Strong secrecy for all message distributions $\Rightarrow$ distinguishing security: Let~$m \in \mathcal{M}_n$ be arbitrary.
If strong secrecy holds for all distributions, then in particular it holds for the
distribution~$p_{m}$ defined by~$p_{m}(m') = 1$ if~$m=m'$ and~$0$ otherwise.
Now, Pinsker's inequality (see~\cite[pp.58-59]{CsiszarKorner81}) asserts that~$\V(p, q)
\leq \sqrt{2   \D(p\Vert q)}$ for any distributions~$p$ and~$q$. We thus have:
\begin{align*}
\V(p_{(\Z^n, m)},&p_{\Z^n}p_{m}) \\
&= \sum_{m'} \int_{\R^n}
\abs{p_{(\Z^n,m)}(\mathbf{z},m')-p_{\Z^n}(\mathbf{z})p_{m}(m')} d\mathbf{z} \\
&=
\int_{\R^n}\abs{p_{\Z^n|\M=m}(\mathbf{z})-p_{\Z^n}(\mathbf{z})}d\mathbf{z}\\
&  \leq \sqrt{2 \I(m;\Z^n)}.
\end{align*}
The strong secrecy assumption implies that:
\begin{equation*}
\V(p_{\Z^n|\M=m},p_{\Z^n}) =\int_{\R^n}\abs{p_{\Z^n|\M=m}(\mathbf{z})-p_{\Z^n}(\mathbf{z})}d\mathbf{z}
\rightarrow 0.
\end{equation*}
Using the triangular inequality
\begin{multline*}
\V(p_{\Z^n|\M=m},p_{\Z^n|\M=m'})\\ \leq \V(p_{\Z^n|\M=m},p_{\Z^n}) +
\V(p_{\Z^n|\M=m'},p_{\Z^n}),
\end{multline*}
we obtain distinguishing security.
\end{IEEEproof}

Note that Lemma 2 in \cite{Csiszar96} also holds: For any
distribution~$q_{\Z^n}$ on $\R^n$, we have
\begin{equation} \label{Lemma2_Csiszar}
d_{\av} \leq 2 \sum_{m \in \mathcal{M}_n} p_{\M}(m) \V(p_{\Z^n|\M=m},q_{\Z^n}).
\end{equation}

Together with Lemma~\ref{Csiszar_lemma}, this leads to an
upper bound on the mutual information, in case we can approximate~$p_{\Z^n|\M=m}$
by a density that is independent of~$m$.
\begin{lem}\label{lemma2}
Suppose that for all $n$ there exists some density~$q_{\Z^n}$ in
$\R^n$ such that $\V(p_{\Z^n|\M=m},q_{\Z^n})\leq \varepsilon_n$, for all
$m \in \mathcal{M}_n$. Then we have $d_{\av}\leq
2\varepsilon_n$ and so
\begin{equation}
\I(\M;\Z^n) \leq 2\varepsilon_n nR -2\varepsilon_n\log (2\varepsilon_n).
\end{equation}

\end{lem}

In the rest of this paper, we will use lattice codes to achieve
semantic security over the wiretap channel.

\section{Lattice Gaussian distribution and flatness factor}

In this section, we introduce the mathematical tools we will need to describe
and analyze our wiretap codes.

\subsection{Preliminaries on Lattices}

An $n$-dimensional {lattice} $\Lambda$ in the Euclidean space
$\mathbb{R}^{n}$ is a set defined by
\begin{equation*}
\Lambda=\mathcal{L}\left( \mathbf{B}\right) =\left\{ \mathbf{Bx}\text{ : }\mathbf{x\in }\text{ }%
\mathbb{Z}^{n}\right\}
\end{equation*}%
where the columns of the basis matrix $\mathbf{B=}\left[
\mathbf{b}_{1}\cdots \mathbf{b}_{n}\right] $ are linearly
independent.
The dual lattice $\Lambda^*$ of a lattice $\Lambda$ is
defined as the set of vectors $\mathbf{v}\in \R^n$ such
that~$\langle\mathbf{v}, \bm{\lambda}\rangle \in
\mathbb{Z}$, for all $\bm{\lambda} \in \Lambda$ (see, e.g., \cite{BK:Conway98}).

For a vector $\mathbf{x}$, the nearest-neighbor quantizer associated
with $\Lambda$ is
$Q_{\Lambda}(\mathbf{x})=\arg\min_{\bm{\lambda} \in\Lambda}\|\bm{\lambda}-\mathbf{x}\|$.
We define the usual modulo lattice operation by $\mathbf{x} \mod \Lambda \triangleq
\mathbf{x} -Q_{\Lambda}(\mathbf{x})$. The Voronoi cell of $\Lambda$,
defined by
$\mathcal{V}(\Lambda)=\{\mathbf{x}:Q_{\Lambda}(\mathbf{x})=\mathbf{0}\}$,
specifies the nearest-neighbor decoding region.
The Voronoi cell is one example of the fundamental region of a lattice. A measurable set~$\mathcal{R}(\Lambda)\subset \mathbb{R}^n$ is a fundamental region of the lattice~$\Lambda$ if~$\cup_{\bm{\lambda} \in \Lambda} (\mathcal{R}(\Lambda)+\bm{\lambda}) = \R^n$ and
if~$(\mathcal{R}(\Lambda)+\bm{\lambda}) \cap (\mathcal{R}(\Lambda)+\bm{\lambda}')$
has measure~$0$ for any~$\bm{\lambda} \neq \bm{\lambda}'$
in~$\Lambda$.
More generally, for a vector $\mathbf{x}$, the mod~$\mathcal{R}(\Lambda)$ operation is defined by
$\mathbf{x}\mapsto \check{\mathbf{x}}$ where $\mathbf{\check{x}}$ is the unique element of
$\mathcal{R}(\Lambda)$ such that $\check{\mathbf{x}} - \mathbf{x} \in \Lambda$.
Obviously, the usual mod-$\Lambda$ operation corresponds to the case where $\mathcal{R}(\Lambda)=\mathcal{V}(\Lambda)$.

For a (full-rank)
sublattice $\Lambda' \subset \Lambda$, the finite
group~$\Lambda/\Lambda'$  is
defined as the group of distinct cosets $\bm{\lambda} + \Lambda'$ for
$\bm{\lambda} \in \Lambda$. Denote by $[\Lambda/\Lambda']$ a set of coset representatives.
The lattices $\Lambda'$ and $\Lambda$ are often said to
form a pair of nested lattices, in which $\Lambda$ is
referred to as the {fine lattice} while $\Lambda'$ the {coarse
lattice}. The order of the quotient group~$\Lambda/\Lambda'$ is equal to $V(\Lambda')/V(\Lambda)$.

We refer the readers to \cite{Poltyrev94, ErezZamir04} for more background on lattice coding, especially the definitions of quantization and AWGN-good lattices.

\subsection{Lattice Theta Series}

The {theta series} of $\Lambda$ (see, e.g., \cite{BK:Conway98}) is defined as
\begin{eqnarray}
\Theta_{\Lambda}(q)=\sum_{\bm{\lambda} \in \Lambda} q^{\|\bm{\lambda}\|^2}
\end{eqnarray}
where $q= e^{j\pi z}$ (imaginary part $\Im(z)>0$). Letting $z$ be purely imaginary, and assuming $\tau=\Im(z)>0$, we can
alternatively express the theta series as
\begin{eqnarray}
\Theta_{\Lambda}(\tau)=\sum_{\bm{\lambda} \in \Lambda} e^{-\pi
\tau\|\bm{\lambda}\|^2}.
\end{eqnarray}

For integer $p>0$, let $\mathbb{Z}^n \rightarrow \mathbb{Z}^n_p:
{\mathbf{v}} \mapsto \overline{\mathbf{v}}$ be the element-wise
reduction modulo-$p$. Following \cite{Loeliger}, consider mod-$p$
lattices (Construction~A) of the form $\Lambda_C \triangleq
\{\mathbf{v}\in \mathbb{Z}^n: \overline{\mathbf{v}} \in C\}$, where
$p$ is a prime and $C$ is a linear code over $\mathbb{Z}_p$.
Equivalently, $\Lambda_C = p\mathbb{Z}^n + C$. In the
proof, scaled mod-$p$ lattices $a \Lambda_C \triangleq \{a
\mathbf{v}: \mathbf{v}\in \Lambda_C\}$ for some $a \in
\mathbb{R}^+$ are used. The fundamental volume of such a lattice is
$V(a \Lambda_C)=a^n p^{n-k}$, where $n$ and $k$ are the
block length and dimension of the code $C$, respectively. A set
$\mathcal{C}$ of linear codes over $\mathbb{Z}_p$ is said to be
balanced if every nonzero element of $\mathbb{Z}_p^n$ is contained in
the same number of codes from~$\mathcal{C}$. In particular, the set
of all linear $(n,k)$ codes over $\mathbb{Z}_p$ is balanced.

\begin{lem}[Average behavior of theta series] \label{theta_average}
Let $\mathcal{C}$ be any balanced set of linear $(n,k)$ codes over
$\mathbb{Z}_p$. Then, for $0<k<n$, for $a^n p^{n-k}=V$
and~$\tau$ fixed, we have:
\begin{equation}\label{theta-average}
\lim_{a\rightarrow0, p\rightarrow
\infty}\frac{1}{|\mathcal{C}|}\sum_{C\in
\mathcal{C}}\Theta_{a\Lambda_C}(\tau) = 1 + \frac{1}{V \tau^{n/2}}.
\end{equation}
\end{lem}

The proof of Lemma~\ref{theta_average} is provided in Appendix~\ref{app:MinkowskiHlawka}.

\subsection{Lattice Gaussian Distribution} \label{lattice_gaussian}

Lattice Gaussian distributions arise from various problems in
mathematics \cite{Banaszczyk}, coding \cite{Forney00} and cryptography \cite{Micciancio05}.
For~$\sigma>0$ and $\mathbf{c} \in \R^n$,
we define the
Gaussian distribution of variance $\sigma^2$ centered at ${\bf c} \in \R^n$ as
\begin{equation*}
 f_{\sigma,{\bf c}}(\mathbf{x})=\frac{1}{(\sqrt{2\pi}\sigma)^n}e^{- \frac{\|\mathbf{x}-{\bf c}\|^2}{2\sigma^2}},
\end{equation*}
for all $\mathbf{x} \in\R^n$. For convenience, we write $f_{\sigma}(\mathbf{x})=f_{\sigma,{\bf
0}}(\mathbf{x})$.

We also consider the $\Lambda$-periodic function
\begin{equation}\label{Guass-function-lattice}
  f_{\sigma,\Lambda}(\mathbf{x})=\sum_{{\bm{\lambda}}\in \Lambda}
{f_{\sigma,{\bm{\lambda}}}(\mathbf{x})}=\frac{1}{(\sqrt{2\pi}\sigma)^n}
\sum_{\bm{\lambda} \in \Lambda} e^{-
    \frac{\|\mathbf{x}-\bm{\lambda}\|^2}{2\sigma^2}},
\end{equation}
for all $\mathbf{x} \in\R^n$. Observe that $f_{\sigma,\Lambda}$ restricted to the quotient $\R^n/\Lambda$ is a probability density.

We define the \emph{discrete Gaussian distribution} over $\Lambda$
centered at $\mathbf{c} \in \R^n$ as the following discrete
distribution taking values in $\bm{\lambda} \in \Lambda$:
\[
D_{\Lambda,\sigma,\mathbf{c}}(\bm{\lambda})=\frac{f_{\sigma,\mathbf{c}}(\mathbf{\bm{\lambda}})}{f_{\sigma,\Lambda}(\mathbf{c})}, \quad \forall \bm{\lambda} \in \Lambda,
\]
since $f_{\sigma,\Lambda}(\mathbf{c}) = \sum_{\bm{\lambda} \in
\Lambda} f_{\sigma,\mathbf{c}}(\mathbf{\bm{\lambda}})$. Again for convenience, we write $D_{\Lambda,\sigma}=D_{\Lambda,\sigma,\mathbf{0}}$.

It will be useful to define the {discrete Gaussian distribution} over a coset of $\Lambda$, i.e., the shifted lattice $\Lambda-\mathbf{c}$:
\[
D_{\Lambda-\mathbf{c},\sigma}(\bm{\lambda}-\mathbf{c})=\frac{f_{\sigma}(\mathbf{\bm{\lambda}}-\mathbf{c})}{f_{\sigma,\Lambda}(\mathbf{c})} \quad \forall \bm{\lambda} \in \Lambda.
\]
Note the relation $D_{\Lambda-\mathbf{c},\sigma}(\bm{\lambda}-\mathbf{c}) = D_{\Lambda,\sigma,\mathbf{c}}(\bm{\lambda})$, namely, they are a shifted version of each other.

\subsection{Flatness Factor}

The flatness factor of a lattice~$\Lambda$ quantifies the maximum variation of~$f_{\sigma,\Lambda}(\mathbf{x})$ for~$\mathbf{x} \in \R^n$.

\begin{deft} [Flatness factor]
For a lattice~$\Lambda$ and for a parameter~$\sigma$, the flatness factor
is defined by:
\begin{equation*}
\epsilon_{\Lambda}(\sigma)  \triangleq \max_{\mathbf{x} \in
\mathcal{R}(\Lambda)}\abs{
V(\Lambda)f_{\sigma,\Lambda}(\mathbf{x})-1}
\end{equation*}
where $\mathcal{R}(\Lambda)$ is a fundamental region of $\Lambda$.
\end{deft}

It is more illustrative to write
\begin{equation*}
\epsilon_{\Lambda}(\sigma) = \max_{\mathbf{x} \in
\mathcal{R}(\Lambda)}\abs{
\frac{f_{\sigma,\Lambda}(\mathbf{x})}{1/V(\Lambda)}-1}.
\end{equation*}
Thus, the flatness factor may be interpreted
as the maximum variation of~$f_{\sigma,\Lambda}(\mathbf{x})$
with respect to the uniform distribution on~$\mathcal{R}(\Lambda)$. In other words, $f_{\sigma,\Lambda}(\mathbf{x})$ is within
$1\pm\epsilon_{\Lambda}(\sigma)$ from the uniform distribution over~$\mathcal{R}(\Lambda)$.
Note that this definition slightly differs from that in~\cite{BelfioreITW11}: The present definition also
takes into account the minimum of~$f_{\sigma,\Lambda}(\mathbf{x})$.

\begin{prop} [Expression of $\epsilon_{\Lambda}(\sigma)$] \label{expression_epsilon}
We have:
\begin{equation*}
\epsilon_{\Lambda}(\sigma) =  \left(\frac{\gamma_{\Lambda}(\sigma)}{{2\pi}}\right)^{\frac{n}{2}}{
\Theta_{\Lambda}\left({\frac{1}{2\pi\sigma^2}}\right)}-1
\end{equation*}
where $\gamma_{\Lambda}(\sigma) = \frac{
V(\Lambda)^{\frac{2}{n}}}{\sigma^2}$ is the volume-to-noise ratio (VNR)\footnote{The definition of VNR varies slightly in literature, by a factor $2\pi$ or $2\pi e$. In particular, the VNR is defined as $V(\Lambda)^{\frac{2}{n}}/(2\pi e\sigma^2)$ in \cite{Forney00,ErezZamir04}, while the generalized signal-to-noise ratio (GSNR) is defined as $V(\Lambda)^{\frac{2}{n}}/(2\pi \sigma^2)$ in the conference version of this paper \cite{LLB12}.}.
\end{prop}

\begin{IEEEproof}
Using the Fourier expansion of
$f_{\sigma,\Lambda}(\mathbf{x})$ over the dual lattice $\Lambda^*$ (see, e.g., \cite{Forney00,Micciancio05}), we
obtain, for all~${\bf x} \in \mathcal{R}(\Lambda)$:
{ \allowdisplaybreaks
\begin{equation*}\label{flatness-proof}
\begin{split}
&\abs{V(\Lambda)f_{\sigma,\Lambda}(\mathbf{x})-1} \\
&= \left|\sum_{{\bf \bm{\lambda}^*}\in \Lambda^*} e^{-2\pi^2\sigma^2\|{\bf \bm{\lambda}^*}\|^2} \cos( 2\pi  \langle{\bf \bm{\lambda}^*},\mathbf{x}\rangle) - 1\right |\\
&\stackrel{(a)}{\leq}
\sum_{{\bf \bm{\lambda}^*}\in \Lambda^*}e^{-2\pi^2 \sigma^2\|{\bf \bm{\lambda}^*}\|^2}-1\\
& \stackrel{(b)}{=} V(\Lambda)f_{\sigma,\Lambda}(\mathbf{0})-1\\
& = \frac{V(\Lambda)}{(\sqrt{2\pi}\sigma)^n}\sum_{{\bf \bm{\lambda}}\in \Lambda}e^{-\frac{\|{\bf \bm{\lambda}}\|^2}{2\sigma^2}}-1\\
& \stackrel{(c)}{=} \frac{V(\Lambda)}{(\sqrt{2\pi}\sigma)^n}\Theta_{\Lambda}\left(\frac{1}{2\pi\sigma^2}\right)-1,
\end{split}
\end{equation*}
}%
where the equality in (a) holds if $\mathbf{x}\in \Lambda$ so that
$\langle{\bf \bm{\lambda}^*},\mathbf{x}\rangle$ is an integer for
all~$\bm{\lambda}^* \in \Lambda^*$, (b) is due
to the Poisson sum formula, and (c) follows from the definition of
the theta series. The result follows.
\end{IEEEproof}

From step (a) of the proof, we can see that:

\begin{corol}
Alternatively, the flatness factor can be expressed on the dual lattice $\Lambda^*$ as
\begin{equation}\label{flatness-dual-lattice}
\epsilon_{\Lambda}(\sigma)=\Theta_{\Lambda^*}\left({{2\pi\sigma^2}}\right)-1.
\end{equation}
\end{corol}

\begin{rem}\label{remark1}
The equality in (a) implies that the maxima of both
$f_{\sigma,\Lambda}(\mathbf{x})$ and
$\abs{f_{\sigma,\Lambda}(\mathbf{x})-1/V(\Lambda)}$ are reached when $\mathbf{x}\in
\Lambda$.
\end{rem}

\begin{rem} \label{monotonicity}
From (\ref{flatness-dual-lattice}),
it is easy to see that $\epsilon_{\Lambda}$ is
a monotonically decreasing function of $\sigma$, i.e., for $\sigma_1 < \sigma_2$,
we have $\epsilon_{\Lambda}(\sigma_2) \leq  \epsilon_{\Lambda}(\sigma_1)$.
\end{rem}

\begin{rem} \label{sublattice-flatness}
If $\Lambda_2$ is a sublattice of $\Lambda_1$, then
$\epsilon_{\Lambda_1}(\sigma) \leq \epsilon_{\Lambda_2}(\sigma)$.
\end{rem}

\begin{rem} \label{invariance-flatness}
The flatness factor is invariant if both $\Lambda$ and $\sigma$ are scaled, i.e., $\epsilon_{\Lambda}(\sigma) = \epsilon_{a\Lambda}(a\sigma)$.
\end{rem}

In the following, we show that the flatness factor is equivalent to the notion
of smoothing parameter\footnote{We remark that
this definition differs slightly from the one in~\cite{Micciancio05}, where~$\sigma$ is scaled by a constant factor $\sqrt{2\pi}$ (i.e., $s=\sqrt{2\pi}\sigma$).} that is commonly used in lattice-based
cryptography.

\begin{deft} [Smoothing parameter \cite{Micciancio05}]
For a lattice $\Lambda$ and for $\varepsilon > 0$, the smoothing parameter
$\eta_{\varepsilon}(\Lambda)$ is the smallest $\sigma>0$ such that
$\sum_{{\bf \bm{\lambda}^*}\in \Lambda^* \setminus \{\mathbf{0}\}} e^{-2\pi^2
\sigma^2\|{\bf \bm{\lambda}^*}\|^2}\leq \varepsilon$.
\end{deft}

\begin{prop}
If $\sigma=\eta_{\varepsilon}(\Lambda)$, then
$\epsilon_{\Lambda}(\sigma) = \varepsilon$.
\end{prop}

\begin{IEEEproof}
From Corollary 1, we can see
that
{\small
\begin{align*}
\epsilon_{\Lambda}(\sigma) =
\sum_{{\bf \bm{\lambda}^*}\in
\Lambda^*}
e^{-2\pi^2 \sigma^2\|{\bf \bm{\lambda}^*}\|^2}-1
= \sum_{{\bf \bm{\lambda}^*}\in \Lambda^* \setminus \{\bf 0\}} e^{-2\pi^2
\sigma^2\|{\bf \bm{\lambda}^*}\|^2} =\varepsilon.
\end{align*}
}
\end{IEEEproof}

Despite the equivalence, the flatness factor has two main technical advantages:

\begin{itemize}
  \item It allows for a direct characterization by the theta
  series, which leads to a much better bound due to Lemma~\ref{theta_average}. Note that it is $\varepsilon$, not the smoothing parameter, that is of more interest to communications.
  \item The studies of the smoothing parameter are mostly concerned with small values of
  $\varepsilon$, while the flatness factor can handle both large and
  small values of $\varepsilon$. This is of interest in
  communication applications~\cite{BelfioreITW11}.
\end{itemize}


\begin{figure*}[ht!]
    \begin{center}
        \subfigure[$\gamma_{\Lambda}(\sigma)=8\pi, \epsilon_{\Lambda}(\sigma) = 3$.]{%
            \label{}
            \includegraphics[width=0.45\textwidth]{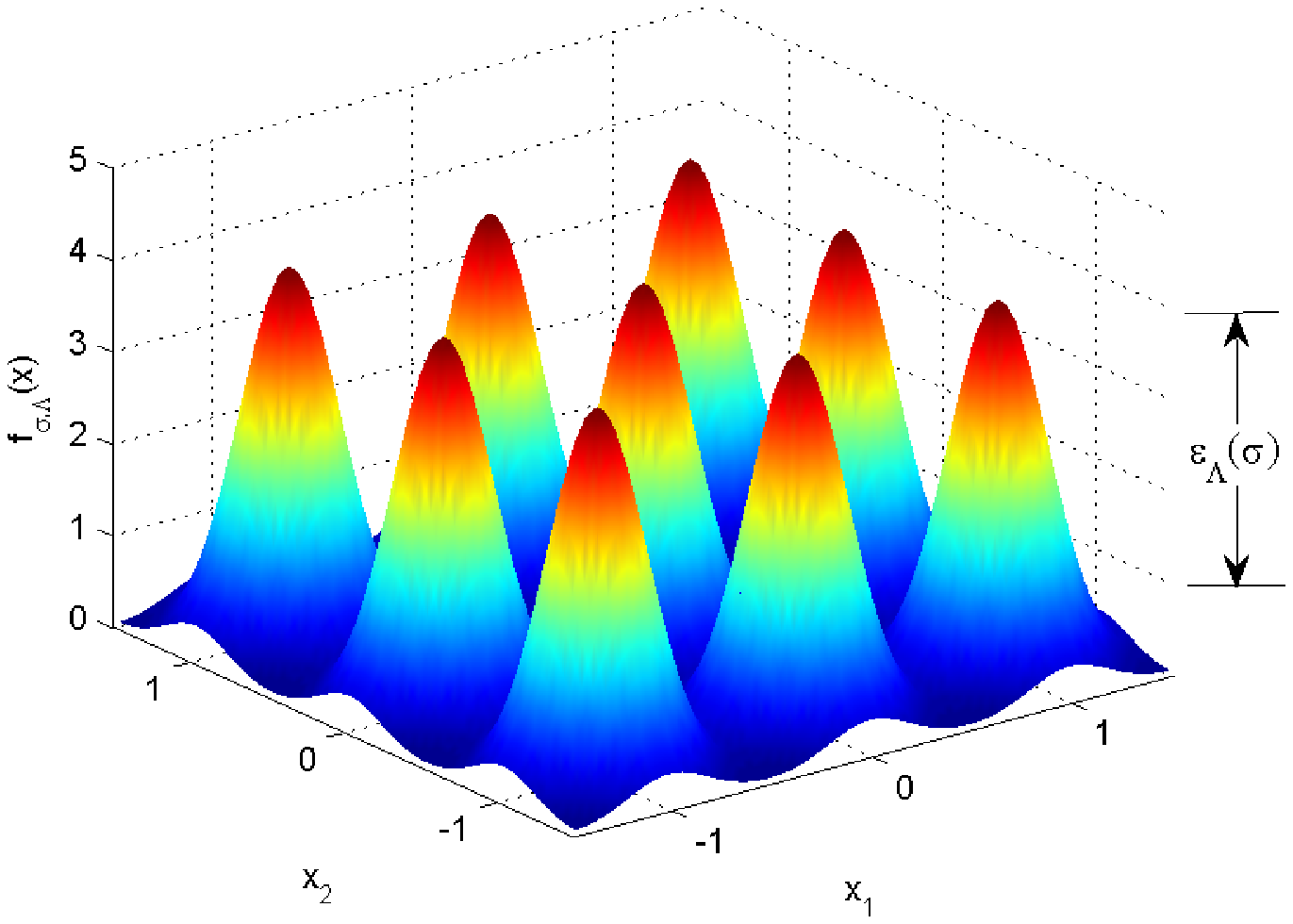}
        }%
        \subfigure[$\gamma_{\Lambda}(\sigma)=\pi, \epsilon_{\Lambda}(\sigma) = 0.0075$.]{%
           \label{}
           \includegraphics[width=0.45\textwidth]{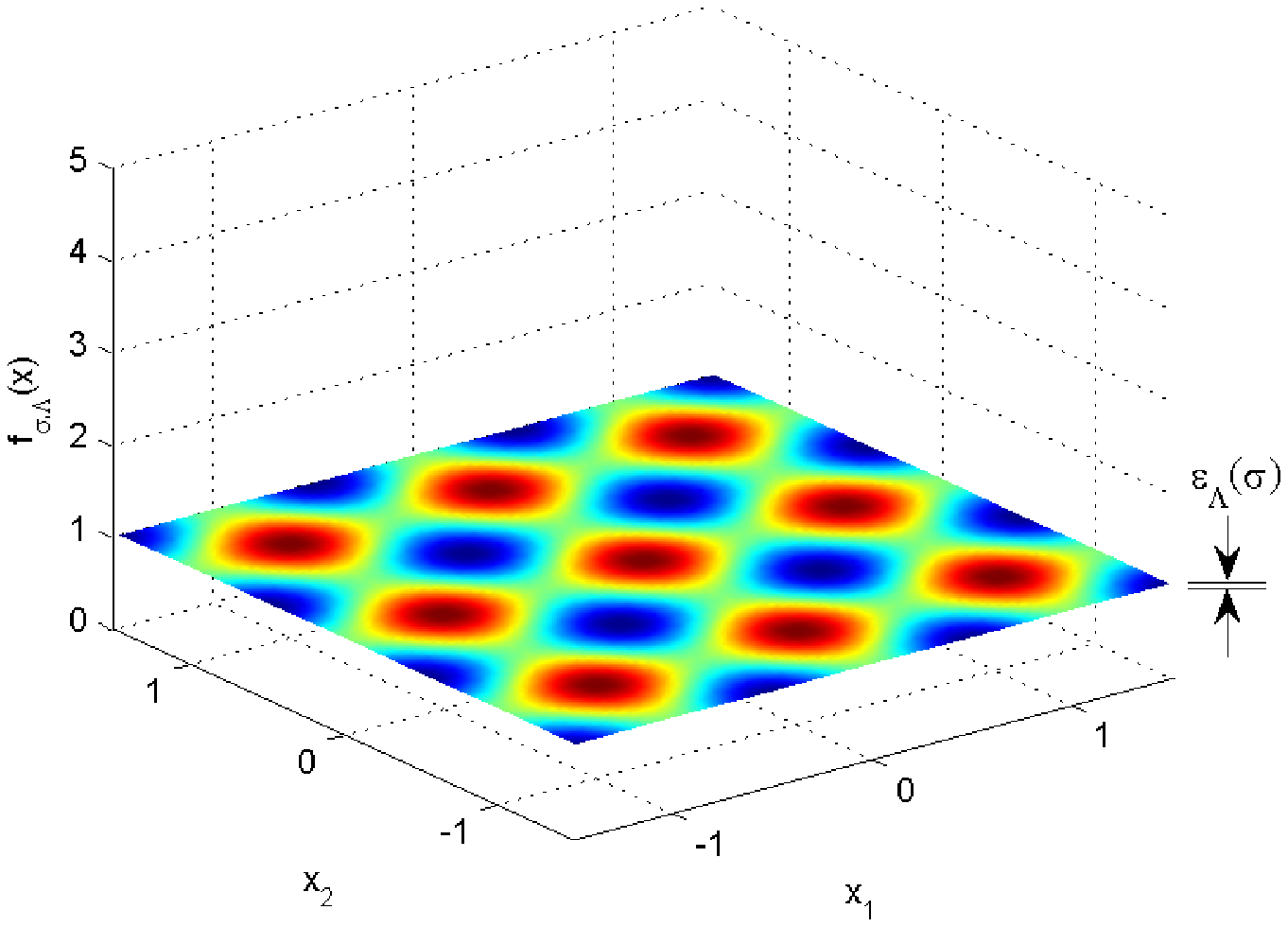}
        }\\ 
%
    \end{center}

    \vspace{-12pt}

    \caption{%
           Lattice Gaussian distribution and flatness factor for $\mathbb{Z}^2$ (a) at high VNR where $\epsilon_{\Lambda}(\sigma)$ is large and the Gaussians are well separated, and (b) at low VNR where $\epsilon_{\Lambda}(\sigma)$ is small and the distribution is nearly uniform.
     }%
    \label{fig:flatness}
\end{figure*}

Figure \ref{fig:flatness} illustrates the flatness factor and lattice Gaussian distribution at different VNRs for lattice $\mathbb{Z}^2$. When the VNR is high (Fig. \ref{fig:flatness}(a)), $\epsilon_{\Lambda}(\sigma)$ is large and the Gaussians are well separated, implying reliable decoding is possible; this scenario is desired in communications. When the VNR is low (Fig. \ref{fig:flatness}(b)), $\epsilon_{\Lambda}(\sigma)$ is small and the distribution is nearly uniform, implying reliable decoding is impossible; this scenario is desired in security and will be pursued in following sections.



The flatness factor also gives a bound on the variational distance
between the Gaussian distribution reduced mod $\mathcal{R}(\Lambda)$
and the uniform distribution $U_{\mathcal{R}(\Lambda)}$ on
$\mathcal{R}(\Lambda)$. This result was proven in
\cite{Micciancio05} using the smoothing parameter when
$\mathcal{R}(\Lambda)$ is the fundamental parallelotope. We give a proof for any $\mathcal{R}(\Lambda)$, for the sake of completeness.

\begin{prop} \label{Micciancio_Regev}
For $\mathbf{c} \in \R^n$, let
$\bar{f}(\cdot)$ be the density function of $\mathbf{x} \Mod \mathcal{R}(\Lambda)$ where $\mathbf{x}\thicksim f_{\sigma,\mathbf{c}}(\cdot)$. Then
$$\V(\bar{f},U_{\mathcal{R}(\Lambda)})\leq \epsilon_{\Lambda}(\sigma).$$
\end{prop}
\begin{IEEEproof}
Observe that restricting $f_{\sigma,\Lambda}$ to any fundamental
region~$\mathcal{R}(\Lambda)$ is equivalent to considering the Gaussian
distribution modulo~$\mathcal{R}(\Lambda)$:
\begin{align*}
\bar{f}(\mathbf{x})& =\sum_{\bm{\lambda} \in \Lambda} f_{\sigma,\mathbf{c}}(\mathbf{x}-\bm{\lambda}) \mathds{1}_{\mathcal{R}(\Lambda)}(\mathbf{x})\\
&=\sum_{\bm{\lambda} \in \Lambda}
f_{\sigma,\bm{\lambda}}(\mathbf{x-c})\mathds{1}_{\mathcal{R}(\Lambda)}(\mathbf{x})\\
&=f_{\sigma,\Lambda}(\mathbf{x-c})\mathds{1}_{\mathcal{R}(\Lambda)}(\mathbf{x}).
\end{align*}
Then by definition of $\epsilon_{\Lambda}(\sigma)$, we find
\begin{align*}
\int_{\mathcal{R}(\Lambda)}&\abs{\bar{f}(\mathbf{t}) - U_{\mathcal{R}(\Lambda)}(\mathbf{t})} d\mathbf{t}  \\
&\leq V(\Lambda) \max_{\mathbf{x} \in
\mathcal{R}(\Lambda)}\abs{f_{\sigma,\Lambda}(\mathbf{x-c}) -
\frac{1}{V(\Lambda)}} \\
& = V(\Lambda) \max_{\mathbf{x} \in
\mathcal{R}(\Lambda)-\mathbf{c}}\abs{f_{\sigma,\Lambda}(\mathbf{x}) -
\frac{1}{V(\Lambda)}}
\leq  \epsilon_{\Lambda}(\sigma), \quad \quad
\quad \quad
\end{align*}
because~$\mathcal{R}(\Lambda)-\mathbf{c}$ is a fundamental region
of~$\Lambda$.
\end{IEEEproof}

By definition, the flatness factor in fact guarantees a stronger property: if
$\epsilon_{\Lambda}(\sigma) \rightarrow 0 $, then
$f_{\sigma,\Lambda}(\mathbf{x})$ converges uniformly to the uniform
distribution on the fundamental region.

The following result guarantees the existence of sequences of lattices whose
flatness factors can respectively vanish or explode
as $n \to \infty$.

\begin{ther}\label{theorem2}
For any $\sigma>0$ and $\delta>0$, there exists a sequence of mod-$p$ lattices~$\Lambda^{(n)}$ such that
\begin{equation} \label{eq:flatness_factor_bound}
\epsilon_{\Lambda^{(n)}}(\sigma) \leq
(1+\delta) \cdot \left(\frac{\gamma_{\Lambda^{(n)}}(\sigma)}{2\pi}\right)^{\frac{n}{2}},
\end{equation}
i.e., the flatness factor goes to zero exponentially for any fixed
VNR (as a function of $n$) $\gamma_{\Lambda^{(n)}}(\sigma)<2\pi$; oppositely,
there also exists a sequence of mod-$p$ lattices $\Lambda'^{(n)}$ such
that
\begin{equation} \label{eq:flatness_factor_bound2}
\epsilon_{\Lambda'^{(n)}}(\sigma) \geq
(1-\delta) \cdot \left(\frac{\gamma_{\Lambda'^{(n)}}(\sigma)}{2\pi}\right)^{\frac{n}{2}},
\end{equation}
i.e., its flatness factor goes to infinity exponentially for any fixed VNR
$\gamma_{\Lambda'^{(n)}}(\sigma)>2\pi$.
\end{ther}

\begin{IEEEproof}
Lemma~\ref{theta_average} guarantees that for all~$n$, $\delta$ and~$\tau$ there
exists~$a(n,\delta,\tau)$ (and the corresponding $p$ such that $a^np^{n-k}=V(\Lambda)$) such that $\mathbb{E}_{C}\left[\Theta_{a\Lambda_{C}}(\tau)\right] \leq 1 + \delta + \frac{1}{V(\Lambda)\tau^{\frac{n}{2}}}$. Here~$C$ is sampled uniformly among all linear~$(n,k)$ codes over~$\mathbb{Z}_p$ and~$a \Lambda_C = \{a {\bf v}: {\bf v} \in \Lambda_C\}$.
Therefore there exists a sequence of lattices~$\Lambda^{(n)}$ such that $\Theta_{\Lambda^{(n)}}(\tau) \leq 1 + \delta + \frac{1}{V(\Lambda^{(n)})\tau^{\frac{n}{2}}}$. For this sequence,
Proposition \ref{expression_epsilon} gives $\epsilon_{\Lambda}(\sigma) \leq (1+\delta) \left(\gamma_{\Lambda}(\sigma)/(2 \pi)\right)^{\frac{n}{2}}$ when we let $\tau=\frac{1}{2\pi\sigma^2}$. The second half of the theorem can be
proved in a similar fashion.
\end{IEEEproof}

Theorem \ref{theorem2} shows a phenomenon of
``phase transition" for the flatness factor, where the boundary is $\gamma_{\Lambda}(\sigma)=2\pi$.

\begin{rem}

In fact, we can show a concentration result on the flatness factor of the ensemble of mod-$p$ lattices, that is,
most mod-$p$ lattices have a flatness factor concentrating around $\left(\gamma_{\Lambda}(\sigma)/(2 \pi)\right)^{\frac{n}{2}}$.
In particular, using the Markov inequality, we see that with probability higher than
$1-2^{-n}$ over the choice of $\Lambda^{(n)}$,
\begin{equation} \label{eq:flatness_factor_concent}
\epsilon_{\Lambda^{(n)}}(\sigma) \leq
(1+\delta) \cdot [2\gamma_{\Lambda^{(n)}}(\sigma)/\pi]^{\frac{n}{2}},
\end{equation}

Thus, for $\gamma_{\Lambda^{(n)}}(\sigma) < \pi/2$, we could have
$\epsilon_{\Lambda}(\sigma)  \to 0$ exponentially.
This is slightly worse than what we have in (12), but it holds with very
high probability, making the construction of the scheme potentially more practical.

\end{rem}

\subsection{Properties of the Flatness Factor}
In this section we collect known properties and further derive new
properties of lattice Gaussian distributions that will be useful in
the paper.

From the definition of the flatness factor and Remark~\ref{remark1}, one can derive
the following result (see also~\cite[Lemma 4.4]{Micciancio05}):

\begin{lem} \label{GVP_Lemma26}
For all~${\bf c} \in \R^n$ and~$\sigma>0$, we have:
\[
\frac{f_{\sigma,{\bf c}}(\Lambda)}{f_{\sigma}(\Lambda)}  \in \left[
\frac{1 - \epsilon_{\Lambda}(\sigma)}{1+\epsilon_{\Lambda}(\sigma)},1\right].
\]
\end{lem}

The following lemma shows that, when the flatness factor of the
coarse lattice is small, a discrete Gaussian distribution over the
fine lattice results in almost uniformly distributed cosets, and
vice versa. The first half of the lemma is a corollary of Lemma~\ref{GVP_Lemma26} (see~\cite[Corollary 2.7]{Gentry08}), while the second
half is proven in Appendix~\ref{Appendix_coset_distribution}.
Let $D_{\Lambda,\sigma,\mathbf{c}}\Mod \Lambda'$ be the short notation for the distribution of
$\mathsf{L} \Mod \Lambda'$ where $\mathsf{L}\thicksim D_{\Lambda,\sigma,\mathbf{c}}$.

\begin{lem} \label{GVP_Lemma27}
Let $\Lambda' \subset \Lambda$ be a pair of nested lattices such that~$\epsilon_{\Lambda'}(\sigma)<\frac{1}{2}$. Then
\[
\V(D_{\Lambda,\sigma,\mathbf{c}}\Mod \Lambda',U(\Lambda/\Lambda')) \leq 4\epsilon_{\Lambda'}(\sigma),
\]
where~$U(\Lambda/\Lambda')$ denotes the uniform distribution over the finite set~$\Lambda/\Lambda'$.
Conversely, if $\mathsf{L}$ is uniformly distributed in $[\Lambda/\Lambda']$ and $\mathsf{L}'$ is sampled from~$D_{\Lambda',\sigma,\mathbf{c}-\mathsf{L}}$, then the distribution $D_{\mathsf{L}+\mathsf{L}'}$ satisfies
\[
\V(D_{\mathsf{L}+\mathsf{L}'},D_{\Lambda,\sigma,\mathbf{c}}) \leq \frac{2\epsilon_{\Lambda'}(\sigma)}{1-\epsilon_{\Lambda'}(\sigma)}.
\]
\end{lem}

The next result shows that the variance per dimension of the
discrete Gaussian $D_{\Lambda,\sigma,\mathbf{c}}$ is not far
from $\sigma^2$ when the flatness factor is small.

\begin{lem} \label{Micciancio_Regev_Lemma43}
Let $\mathsf{L}$ be sampled from the
Gaussian distribution $D_{\Lambda,\sigma,\mathbf{c}}$. If
$\varepsilon \triangleq \epsilon_{\Lambda}\left(\sigma/\sqrt{\frac{\pi}{\pi-1/e}}\right) < 1$, then
\begin{equation}\label{eq:lem6}
\abs{\mathbb{E}\left[\norm{\mathsf{L}-\mathbf{c}}^2\right]-n\sigma^2} \leq \frac{2\pi\varepsilon}{1-\varepsilon}\sigma^2.
\end{equation}
\end{lem}

Lemma \ref{Micciancio_Regev_Lemma43} slightly improves upon Lemma 4.3 in \cite{Micciancio05}, which had another factor $n$ on the right-hand side, and also required $\epsilon_{\Lambda}\left(\sigma/2\right) < 1$. The details are given in Appendix \ref{Appendix_Lemma43}.

\begin{rem}
Note that the coefficient $\sqrt{\frac{\pi}{\pi-1/e}}\approx 1.06$. As shown in Appendix \ref{Appendix_Lemma43}, it is possible to replace the condition
$ \epsilon_{\Lambda}\left(\sigma/1.06\right) < 1$ by
$ \epsilon_{\Lambda}\left(\sigma/c\right) < 1$, where $c$ is arbitrarily close to 1 (but there is another constant $C$ on the right-hand side of (\ref{eq:lem6}) which grows when $c$ tends to 1).
\end{rem}

From the maximum-entropy principle \cite[Chap.~11]{BK:Cover}, it
follows that the discrete Gaussian distribution maximizes the
entropy given the average energy and given the same support over a lattice.
The following lemma further shows that if the flatness factor is
small, the entropy of a discrete Gaussian
$D_{\Lambda,\sigma,\mathbf{c}}$ is almost equal to the differential
entropy of a continuous Gaussian vector of variance~$\sigma^2$ per dimension, minus $\log V(\Lambda)$,
that of a uniform distribution over the fundamental region of $\Lambda$.

\begin{lem}[Entropy of discrete Gaussian] \label{proposition4}
Let $\mathsf{L} \sim
D_{\Lambda,\sigma,\mathbf{c}}$. If
$\varepsilon \triangleq \epsilon_{\Lambda}\left(\sigma/\sqrt{\frac{\pi}{\pi-1/e}}\right)<1$,
then the entropy of~$\mathsf{L}$ satisfies
\[
\abs{\mathbb{H}(\mathsf{L}) - \left[n\log (\sqrt{2 \pi e}\sigma) - \log {V(\Lambda)}\right]} \leq \varepsilon',
\]
where $\varepsilon'= - {\log(1-\varepsilon)} + \frac{\pi\varepsilon}{1-\varepsilon}$.
\end{lem}

\begin{IEEEproof}
By using the identity~$f_{\sigma,{\bf c}}(\bm{\lambda}) = \frac{1}{(\sqrt{2\pi}\sigma)^n}e^{-\frac{\norm{\bm{\lambda}-{\bf c}}^2}{2\sigma^2}}$, we obtain:
 {\allowdisplaybreaks
\begin{align*}
 \mathbb{H}(\mathsf{L})
&=- \sum_{\bm{\lambda} \in \Lambda} \frac{f_{\sigma, {\bf c}}(\bm{\lambda})}{f_{\sigma,\Lambda}(\mathbf{c})}  \log\left(\frac{f_{\sigma, {\bf c}}(\bm{\lambda})}{f_{\sigma,\Lambda}(\mathbf{c})}\right)\\
&= \log \left((\sqrt{2\pi}\sigma)^n f_{\sigma,{\bf c}}(\Lambda)\right)+ \sum_{\bm{\lambda} \in \Lambda} \frac{f_{\sigma, {\bf c}}(\bm{\lambda})}{f_{\sigma,\Lambda}(\mathbf{c})}  \frac{\norm{\bm{\lambda}-\mathbf{c}}^2}{2\sigma^2}\\
&=\log \left((\sqrt{2\pi}\sigma)^n f_{\sigma,{\bf c}}(\Lambda)\right) + \frac{1}{2\sigma^2}\mathbb{E} \left[ \norm{\mathsf{L}-\mathbf{c}}^2\right].
\end{align*}
}%
Due to the definition of the flatness factor, we have
\[
f_{\sigma,\Lambda}(\mathbf{c}) \in \left[\frac{1-\epsilon_{\Lambda}(\sigma)}{V(\Lambda)}, \frac{1+\epsilon_{\Lambda}(\sigma)}{V(\Lambda)}\right].
\]
Moreover, Lemma \ref{Micciancio_Regev_Lemma43} implies
\[
\frac{1}{2\sigma^2} \mathbb{E}\left[\norm{\mathsf{L}-\mathbf{c}}^2\right]\in \left[ \frac{n}{2} -\frac{\pi\varepsilon}{1-\varepsilon},\frac{n}{2} +\frac{\pi\varepsilon}{1-\varepsilon}\right].
\]
Since $\epsilon_{\Lambda}(\sigma) < \epsilon_{\Lambda}(\sigma/2) = \varepsilon$, we have
\begin{multline*}
\abs{\mathbb{H}(\mathsf{L}) -\left[n\log(\sqrt{2\pi e}\sigma) - \log V(\Lambda)\right]} \\ <\max \left\{\log(1+\varepsilon) , -\log(1-\varepsilon) \right\} + \frac{\pi \varepsilon}{1-\varepsilon}.
\end{multline*}
The proof is completed.%
\end{IEEEproof}


The following lemma by Regev (adapted from~\cite[Claim~3.9]{Re09})
shows that if the flatness factor is small, the sum of a discrete
Gaussian and a continuous Gaussian is very close to a continuous
Gaussian.
\begin{lem} \label{Regev_Claim39}
Let $\mathbf{c} \in \R^n$ be
any vector, and $\sigma_0,\sigma>0$. Consider the continuous
distribution~$g$
on~$\R^n$ obtained by adding a continuous Gaussian
of variance $\sigma^2$ to a discrete
Gaussian~$D_{\Lambda-\mathbf{c},\sigma_0}$:
\[
g(\mathbf{x})=\frac{1}{f_{\sigma,\Lambda}(\mathbf{c}) } \sum_{\mathbf{t} \in \Lambda -\mathbf{c}} f_{\sigma_0}(\mathbf{t}) f_{\sigma}(\mathbf{x}-\mathbf{t}).
\]
If~$\varepsilon \triangleq
\epsilon_{\Lambda}\left(\frac{\sigma_0\sigma}{\sqrt{\sigma_0^2+\sigma^2}}\right)
<\frac{1}{2}$,
then $\frac{g(\mathbf{x})}{f_{\sqrt{\sigma_0^2+\sigma^2}}(\mathbf{x})}$ is uniformly close to $1$:
\begin{equation} \label{Regev_uniform_convergence}
\forall \mathbf{x} \in \R^n, \quad
\abs{\frac{g(\mathbf{x})}{f_{\sqrt{\sigma_0^2+\sigma^2}}(\mathbf{x})}-1}
\leq 4 \varepsilon.
\end{equation}
In particular, the distribution $g(\mathbf{x})$ is close to the continuous Gaussian density $f_{\sqrt{\sigma_0^2+\sigma^2}}$ in $L^1$ distance:
\[
\V\left(g,f_{\sqrt{\sigma_0^2+\sigma^2}}\right)\leq 4\varepsilon.
\]
\end{lem}

\section{Mod-$\Lambda$ Gaussian Wiretap Channel} \label{mod_Lambda_section}

Before considering the Gaussian wiretap channel, we will tackle a
simpler model where a modulo lattice operation is performed at both
the legitimate receiver's and eavesdropper's end. That is, both the
legitimate channel and the eavesdropper's channel are mod-$\Lambda$
channels. The mod-$\Lambda$ channel is more tractable and captures
the essence of the technique based on the flatness factor.

\subsection{Channel Model} \label{mod_Lambda_channel_model}

Let $\Lambda_s \subset \Lambda_e \subset \Lambda_b$ be a nested
chain of $n$-dimensional lattices in $\R^n$ such that
\[
\frac{1}{n}\log \abs{\Lambda_b/\Lambda_e}=R, \quad \frac{1}{n} \log \abs{\Lambda_e/\Lambda_s}=R'.
\]
We consider the mod-$\Lambda_s$ wiretap channel depicted in Figure~\ref{fig:modLambdachannel}. The input $\X^n$ belongs to the
Voronoi region $\mathcal{V}(\Lambda_s)$ (i.e., $\Lambda_s$ is the
shaping lattice), while the outputs $\Y^n$ and $\Z^n$ at Bob and
Eve's end respectively are given by
\begin{equation} \label{mod_lambda_wiretap}
\begin{cases}
\Y^n=[\X^n + \mathsf{W}_b^n] \Mod \Lambda_s,\\
\Z^n=[\X^n + \mathsf{W}_e^n] \Mod \Lambda_s,
\end{cases}
\end{equation}
where $\mathsf{W}_b^n$, $\mathsf{W}_e^n$ are $n$-dimensional
Gaussian vectors with zero mean and variance $\sigma_b^2$,
$\sigma_e^2$ respectively.

As in the classical Gaussian channel, the transmitted codebook
$\mathcal{C}$ must satisfy the average power constraint
(\ref{power_constraint}). We denote this wiretap channel by
$W(\Lambda_s,\sigma_b,\sigma_e,P)$. Let $\SNR_b=P/\sigma_b^2$ and $\SNR_e=P/\sigma_e^2$ be the signal-to-noise ratios (SNR) of Bob and Eve, respectively.

\begin{center}
\begin{figure}[hbt]
\begin{footnotesize}
\begin{tikzpicture}[
nodetype1/.style={
    rectangle,
    rounded corners,
    minimum width=12mm,
    minimum height=7mm,
    dashed,
    draw=black
},
nodetype2/.style={
    rectangle,
    rounded corners,
    minimum width=8mm,
    minimum height=7mm,
    draw=black
},
tip2/.style={-latex,shorten >=0.4mm}
]
\matrix[row sep=1.2cm, column sep=0.7cm, ampersand replacement=\&]{
\node (Alice) {\hspace*{-2mm}\textsc{Alice}};  \& \node (encoder) [nodetype2]   {\textsc{enc.}}; \&
\node (W) {$\bigoplus$}; \& \node (modLambda1) [nodetype2] {$\Mod \Lambda_s$}; \&
\node (decoder) [nodetype2] {\textsc{dec.}}; \&
\node (Bob) {\textsc{Bob \hspace*{-2mm}}};\\
\& (invisible) \&
\node (We)  {$\bigoplus$}; \& \node (modLambda2) [nodetype2] {$\Mod \Lambda_s$}; \&
\node (Eve) {\textsc{Eve}}; \&  \\};
\draw[->] (Alice) edge[tip2] node [above] {$\M,\mathsf{S}$} (encoder) ;
\draw[->] (encoder) edge[tip2] node [above] (X) {$\X^n$} (W) ;
\draw[->] (W) edge[tip2] node [above] {$\bar{\Y}^n$} (modLambda1) ;
\draw[->] (modLambda1) edge[tip2] node [above] {$\Y^n$} (decoder) ;
\draw[->] (decoder) edge[tip2] node [above] {$\hat{\M}$} (Bob) ;
\draw[->] (We) edge[tip2] node [above] {$\bar{\Z}^n$} (modLambda2) ;
\draw[->] (modLambda2) edge[tip2] node [above] {$\Z^n$} (Eve) ;
\draw[->,>=latex] (X) |- node [anchor=east] {} (We);
\node[above=0.5cm] (Nb) at (W.north) {$\mathsf{W}_b^n$};
\draw[->] (Nb) edge[tip2] (W);
\node[below=0.5cm] (Ne) at (We.south) {$\mathsf{W}_e^n$};
\draw[->] (Ne) edge[tip2] (We);
\end{tikzpicture}
\end{footnotesize}
\caption{The mod-$\Lambda_s$ Gaussian wiretap channel.}
\label{fig:modLambdachannel}
\end{figure}
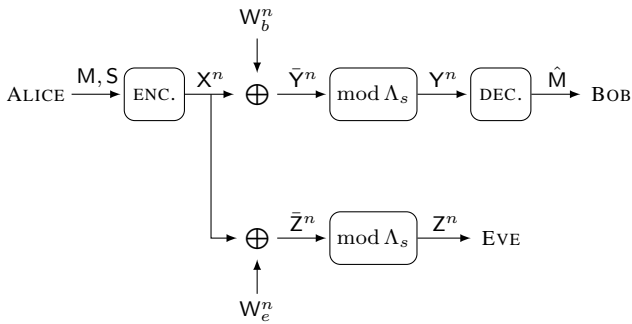
\end{center}

\begin{rem} \label{mod_Lambda_capacity}
As was shown in \cite{Forney00}, the capacity of a mod-$\Lambda$
channel (without MMSE filtering)\footnote{It is known that if an MMSE filter is
added before the mod-$\Lambda$ operation, there exists a sequence of
lattices approaching the capacity of the AWGN channel \cite{ErezZamir04,Forney_MMSE}. However, MMSE filtering is not considered in this section.}
with noise variance $\sigma^2$
is achieved by the uniform distribution on $\mathcal{V}(\Lambda)$
and is given by
\begin{equation} \label{mod_Lambda_capacity}
C(\Lambda,\sigma^2)=\frac{1}{n}\left(\log(V(\Lambda))-h(\Lambda,\sigma^2)\right),
\end{equation}
where $h(\Lambda,\sigma^2)$ is the differential entropy of the
$\Lambda$-aliased noise $\bar{\mathsf{W}}^n=[\mathsf{W}^n] \Mod
\Lambda$. Intuitively, the shaping lattice $\Lambda_s$ must have a
big flatness factor for Bob, otherwise $\bar{\mathsf{W}}^n$ will
tend to a uniform distribution such that the capacity is small.

However, to the best of our knowledge, determining the secrecy capacity of the mod-$\Lambda$ \emph{wiretap} channel (\ref{mod_lambda_wiretap}) is still an open problem. Corollary 2 in \cite{Bloch_Laneman} provides the lower bound
$$C_s \geq C(\Lambda_s,\sigma_b^2)-C(\Lambda_s,\sigma_e^2).$$
\end{rem}

\subsection{Nested Lattice Codes for Binning} \label{mod_Lambda_coding_scheme}
Consider a message set $\mathcal{M}_n=\{1,\ldots,e^{nR}\}$, and a one-to-one function $f: \mathcal{M}_n \to \Lambda_b/\Lambda_e$ which associates each message $m \in \mathcal{M}_n$ to a coset $\widetilde{\bm{\lambda}}_m \in \Lambda_b / \Lambda_e$. We make no \textit{a priori} assumption on the distribution of~$m$. Also, the set of coset representatives $\{\bm{\lambda}_m\}$ are not unique: One could choose $\bm{\lambda}_m \in \Lambda_b \cap \mathcal{R}(\Lambda_e)$ for any fundamental region $\mathcal{R}(\Lambda_e)$, not necessarily the Voronoi region $\mathcal{V}(\Lambda_e)$.

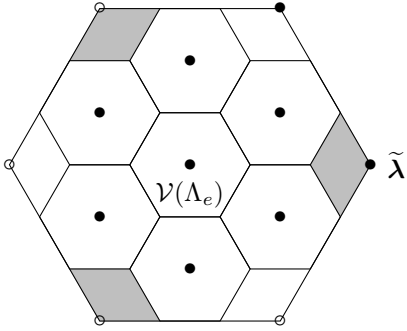
\begin{figure}
\begin{center}
\begin{tikzpicture}[scale=0.8,x={(1,0)},y={(0.5,0.5*sqrt(3))}]
\filldraw[color=lightgray] (3,0) -- (2,1) -- (2,0) -- (3,-1) -- cycle;
\filldraw[color=lightgray] (-3,3) -- (-3,2) -- (-2,2) -- (-2,3) -- cycle;
\filldraw[color=lightgray] (0,-2) -- (-1,-2) -- (0,-3) -- (1,-3) -- cycle;

\draw (3,0) --(0,3) -- (-3,3)--(-3,0)--(0,-3) -- (3,-3) --cycle ;
\foreach \x/\y in {0/0,1/1,-1/2,-1/-1,1/-2,-2/1,2/-1}
{
\draw[shift={(\x,\y)}] node (0,0) {$\bullet$};
\draw[shift={(\x,\y)}] (1,0) --(0,1) -- (-1,1)--(-1,0)--(0,-1) -- (1,-1) --cycle ;
}

\path (0,0) node[below=0.1cm]{$\mathcal{V}(\Lambda_e)$};
\path (3,0) node[right=0.1cm]{$\widetilde{\bm{\lambda}}$};
\node at (3,0) {$\bullet$};
\node at (-3,0) {$\circ$};
\node at (0,3) {$\bullet$};
\node at (0,-3) {$\circ$};
\node at (3,-3) {$\circ$};
\node at (-3,3) {$\circ$};
\end{tikzpicture}
\end{center}

\caption{The grey area represents the region $\mathcal{R}(\widetilde{\bm{\lambda}})$ defined in (\ref{region}) for the lattice pair $\Lambda_e=A_2$, $\Lambda_s=3A_2$, with $\widetilde{\bm{\lambda}}=(3,0)$.}
\label{shifted_Voronoi_regions}
\end{figure}

In order to encode the message $m$, Alice selects a random lattice
point $\bm{\lambda} \in \Lambda_e \cap \mathcal{V}(\Lambda_s)$ according
to the discrete uniform distribution
$p_{\mathsf{L}}(\bm{\lambda})=\frac{1}{e^{nR'}}$ and transmits
$\X^n=\bm{\lambda}+\bm{\lambda}_m$. For $\widetilde{\bm{\lambda}} \in \Lambda_e/\Lambda_s$, define
\begin{align*}
\mathcal{R}(\widetilde{\bm{\lambda}})& =\left(\mathcal{V}(\Lambda_e)+\widetilde{\bm{\lambda}}\right) \Mod \Lambda_s \\
& =\sum_{\bm{\lambda}_s \in \Lambda_{s}} \left(\mathcal{V}(\Lambda_e)+\widetilde{\bm{\lambda}}+\bm{\lambda}_s\right) \cap \mathcal{V}(\Lambda_s).
\end{align*}
The $\mathcal{R}(\widetilde{\bm{\lambda}})$'s are fundamental regions of $\Lambda_e$ and
\begin{equation} \label{region}
\bigcup_{\widetilde{\bm{\lambda}} \in \Lambda_s/\Lambda_e} \mathcal{R}(\widetilde{\bm{\lambda}})=\mathcal{V}(\Lambda_s).
\end{equation}
Figure \ref{shifted_Voronoi_regions} illustrates this relation by an example where $\Lambda_e=A_2$ and $\Lambda_s=3A_2$.

To satisfy the power constraint, we choose a shaping lattice whose second moment per dimension $\sigma^2(\Lambda_s^{(n)})=P$. Under the continuous approximation for large constellations (which could further be made precise by applying a dither), the transmission power will be equal to $P$.

\subsection{A Sufficient Condition for Strong Secrecy} \label{mod_Lambda_secrecy}
We now apply the continuous version of Csisz\`ar's Lemma (Lemma~\ref{Csiszar_lemma}) to derive
an upper bound on the amount of leaked information on the
mod-$\Lambda_s$ wiretap channel (\ref{mod_lambda_wiretap}). Note
that even though we consider a mod-$\Lambda_s$ channel, the secrecy
condition is given in terms of the flatness factor of the lattice~$\Lambda_e$.

\begin{ther}\label{theorem1}
Suppose that the flatness factor of $\Lambda_e$ is
$\varepsilon_n \triangleq \epsilon_{\Lambda_e}(\sigma_e)$ on the eavesdropper's channel. Then
\begin{equation}\label{eq:leakage-bound-mod-Lambda}
\I(\M;\Z^n) \leq 2\varepsilon_n nR
-2\varepsilon_n \log (2 \varepsilon_n).
\end{equation}
\end{ther}

\begin{IEEEproof}
Let $\bar{\Z}^n=\X^n + \mathsf{W}_e^n$. We have, for any message~$m$:
\begin{align*}
p_{\bar{\Z}^n|\M=m}(\mathbf{z})&=\sum_{\bm{\lambda} \in \Lambda_e \cap \mathcal{V}(\Lambda_s)} p_{\mathsf{L}}(\bm{\lambda}) \cdot p_{\bar{\Z}^n|\X^n}(\mathbf{z}|\bm{\lambda}+\bm{\lambda}_m)\\
&=\frac{1}{e^{nR'}} \sum_{\bm{\lambda} \in \Lambda_e \cap \mathcal{V}(\Lambda_s)}
f_{\sigma_e, \bm{\lambda}_m+\bm{\lambda}}({\bf z}).
\end{align*}
The output distribution of Eve's channel conditioned on $m$ having been sent
is then given by
{\allowdisplaybreaks
\begin{align*} \label{conditional_output}
p_{\Z^n|\M=m}(\mathbf{z})&=p_{(\bar{\Z}^n \Mod \Lambda_s)|\M=m}(\mathbf{z})\\
&= \frac{1}{e^{nR'}} \sum_{\bm{\lambda} \in \Lambda_e} \mathds{1}_{\mathcal{V}(\Lambda_s)}(\mathbf{z}) \cdot
f_{\sigma_e, \bm{\lambda}_m+\bm{\lambda}}({\bf z}) \\
&= \frac{1}{e^{nR'}} \sum_{\widetilde{\bm{\lambda}} \in \Lambda_e/\Lambda_s}
 \sum_{\bm{\lambda} \in \Lambda_e} \mathds{1}_{\mathcal{R}(\widetilde{\bm{\lambda}})}(\mathbf{z})
\cdot f_{\sigma_e, \bm{\lambda}_m+\bm{\lambda}} ({\bf z}) \\
&=\frac{1}{e^{nR'}} \sum_{\widetilde{\bm{\lambda}} \in \Lambda_e/\Lambda_s}
  \sum_{\bm{\lambda} \in \Lambda_e}  \mathds{1}_{\mathcal{R}(\widetilde{\bm{\lambda}})}(\mathbf{z}) \cdot f_{\sigma_e,\bm{\lambda}_m}(\mathbf{z}-\bm{\lambda})\\
 &= \frac{1}{e^{nR'}} \sum_{\widetilde{\bm{\lambda}} \in \Lambda_e/\Lambda_s}  \bar{f}_{\widetilde{\bm{\lambda}}}(\mathbf{z}),
\end{align*}
where
$\bar{f}_{\widetilde{\bm{\lambda}}}(\mathbf{z})= \sum_{\bm{\lambda} \in \Lambda_e}  \mathds{1}_{\mathcal{R}(\widetilde{\bm{\lambda}})}(\mathbf{z}) \cdot f_{\sigma_e,\bm{\lambda}_m}(\mathbf{z}-\bm{\lambda})$
is the density function of a continuous Gaussian with variance~$\sigma_e^2$ and center~$\bm{\lambda}_m$ reduced modulo the fundamental region~$\mathcal{R}(\widetilde{\bm{\lambda}})$.
From Proposition \ref{Micciancio_Regev}, we have that  $\V(\bar{f}_{\widetilde{\bm{\lambda}}},U_{\mathcal{R}(\widetilde{\bm{\lambda}})})
\leq \epsilon_{\Lambda_e}(\sigma_e)$ for all $\widetilde{\bm{\lambda}} \in
\Lambda_e/\Lambda_s$. From the decomposition
$U_{\mathcal{V}(\Lambda_s)}(\mathbf{z})=\frac{1}{e^{nR'}} \sum_{\widetilde{\bm{\lambda}} \in
\Lambda_e/\Lambda_s}
 U_{\mathcal{R}(\widetilde{\bm{\lambda}})}(\mathbf{z})$,
we obtain
\begin{align*}
\V(p_{\Z^n|\M=m}&,U_{\mathcal{V}(\Lambda_s)}) \\
& \leq \frac{1}{e^{nR'}} \sum_{\widetilde{\bm{\lambda}} \in
\Lambda_e/\Lambda_s}
\int_{\mathcal{R}(\widetilde{\bm{\lambda}})}
\abs{\bar{f}_{\widetilde{\bm{\lambda}}}(\mathbf{z})-U_{\mathcal{R}(\widetilde{\bm{\lambda}})}(\mathbf{z})}
d\mathbf{z} \\
& \leq \epsilon_{\Lambda_e}(\sigma_e).
\end{align*}}

Recalling the definition of $d_{\av}$ in Lemma
\ref{Csiszar_lemma}, defining
$q_{\Z}(\mathbf{z})=U_{\mathcal{V}(\Lambda_s)}(\mathbf{z})$, and using
the inequality (\ref{Lemma2_Csiszar}),  we find that $d_{\av} \leq
2\epsilon_{\Lambda_e^{(n)}}(\sigma_e)$. Then the mutual information
can be estimated using Lemma \ref{lemma2}.
\end{IEEEproof}

From Theorem \ref{theorem1}, we obtain a sufficient condition for a
sequence of nested lattice wiretap codes to achieve strong secrecy.

\begin{corol}
For any sequence of lattices
$\Lambda_e^{(n)}$ such that
$\epsilon_{\Lambda_e^{(n)}}(\sigma_e)=o\left(\frac{1}{n}\right)$ as $n \to \infty$, we have~$\I(\M;\Z^n) \to 0$.
\end{corol}

In fact, Theorem \ref{theorem2} guarantees the existence of mod-$p$ lattices $\Lambda_e^{(n)}$
whose flatness factor is exponentially small. Therefore, if Eve's
generalized SNR $\gamma_{\Lambda_e}(\sigma_e)$ is smaller than 1,
then strong secrecy can be achieved by such lattice codes, and in that setup
the mutual information will vanish exponentially fast.

Now, we introduce the notion of secrecy-good lattices. Roughly speaking, a lattice is good for secrecy
if its flatness factor is small. Although $\epsilon_{\Lambda_e^{(n)}}(\sigma_e)=o\left(\frac{1}{n}\right)$
is sufficient to achieve strong secrecy, it is desired in practice that the information leakage is exponentially small.
Thus, we define secrecy-goodness as follows:

\begin{deft}[Secrecy-good lattices] \label{secrecy_goodness}
A sequence of lattices $\Lambda^{(n)}$ is \emph{secrecy-good} if
\begin{equation} \label{eq:secrecy-good}
\epsilon_{\Lambda^{(n)}}(\sigma) = e^{-\Omega(n)}, \quad \forall \gamma_{\Lambda^{(n)}}(\sigma)<2\pi.
\end{equation}
\end{deft}

This definition is slightly more general than (\ref{eq:flatness_factor_bound}) of Theorem \ref{theorem2}. The purpose is to accommodate the lattices whose theta series are close to, but not strictly below the Minkowski-Hlawka bound.

\subsection{Existence of Good Wiretap Codes from Nested Lattices}

A priori, the secrecy-goodness property established in the previous
subsection may come at the expense of reliability for the legitimate
receiver. We will show that this is not the case, i.e., that there
exists a sequence of nested lattices which guarantee both strong
secrecy rates and reliability:

\begin{prop} \label{existence_prop}
Given $R, R'>0$, there exists a sequence of
nested lattices $\Lambda_s^{(n)} \subset \Lambda_e^{(n)} \subset
\Lambda_b^{(n)}$ whose nesting ratios satisfy
$$R'_n=\frac{1}{n} \log \frac{V(\Lambda_s)}{V(\Lambda_e)} \to R', \quad R_n=\frac{1}{n} \log \frac{V(\Lambda_e)}{V(\Lambda_b)} \to R$$
when $n \to \infty$, and such that
\begin{enumerate}
\item[-] $\Lambda_s^{(n)}$ is quantization and AWGN-good,
\item[-] $\Lambda_e^{(n)}$ is secrecy-good,
\item[-] $\Lambda_b^{(n)}$ is AWGN-good.
\end{enumerate}
\end{prop}

The proof of Proposition \ref{existence_prop} can be found in Appendix \ref{existence_prop_proof} and follows the approach of \cite{Erez_Litsyn_Zamir}. The main novelty is the addition of the secrecy-goodness property, which requires checking that the corresponding condition is compatible with the ones introduced in \cite{Erez_Litsyn_Zamir}.

\begin{ther} \label{rates_mod_Lambda_prop}
Let $\sigma_e^2>{e}\cdot \sigma^2_b$. Then as $n\rightarrow\infty$, all
strong secrecy rates~$R$ satisfying
\[
R< \frac{1}{2}\log \frac{\sigma_e^2}{\sigma_b^2} -\frac{1}{2}
\]
are achievable using nested lattice codes $\Lambda^{(n)}_s
\subset \Lambda^{(n)}_e \subset \Lambda^{(n)}_b$ on the
mod-$\Lambda^{(n)}_s$ wiretap channel
$W(\Lambda_s,\sigma_b,\sigma_e,P)$.
\end{ther}

\begin{IEEEproof}
Consider the binning scheme described in Section~\ref{mod_Lambda_coding_scheme}, where the nested lattices
$\Lambda^{(n)}_s \subset \Lambda^{(n)}_e \subset \Lambda^{(n)}_b$
are given by Proposition \ref{existence_prop}. Since
$\Lambda^{(n)}_b$ is AWGN-good, without MMSE
filtering, a channel coding rate (without secrecy
constraint) $R+R'<\frac{1}{2} \log \SNR_b$ is
achievable at the legitimate receiver's end, with the error probability vanishing exponentially fast in $n$ \cite{ErezZamir04}.

Since $\Lambda^{(n)}_e$ is secrecy-good, by Theorem \ref{theorem2}
in order to have strong secrecy at the
eavesdropper's end, it is sufficient for mod-$p$ lattices to have
\[
\gamma_{\Lambda_e}(\sigma_e)=\frac{V(\Lambda_s)^{\frac{2}{n}}}{(e^{nR'})^{\frac{2}{n}}\sigma_e^2}
\rightarrow\frac{P\cdot 2\pi e}{e^{2R'}\sigma_e^2}<2\pi,
\]
where $V(\Lambda_s)^{\frac{2}{n}} \rightarrow2\pi e
\sigma^2(\Lambda_s^{(n)})$ because $\Lambda_s^{(n)}$ is
quantization-good and also $P = \sigma^2(\Lambda_s^{(n)})$ under the continuous approximation. The above relation implies
\begin{equation} \label{eq:binrate}
R'>\frac{1}{2}\log\SNR_e+\frac{1}{2}.
\end{equation}
Consequently, all strong secrecy
rates~$R$ satisfying
\[
R< \frac{1}{2}\log \frac{\sigma_e^2}{\sigma_b^2} -\frac{1}{2}
\]
are achievable on the wiretap channel $W(\Lambda_s,\sigma_b,\sigma_e,P)$. Note that positive rates are achievable by the proposed scheme only if $\sigma_e^2>{e} \cdot \sigma_b^2$.
\end{IEEEproof}

For high SNR, the strong secrecy rate that can be achieved using Proposition~\ref{rates_mod_Lambda_prop} is very
close to the lower bound on the secrecy capacity, to within a half nat.

\begin{rem} \label{resolvability_remark}
In our strong secrecy scheme, the output distribution of each bin with respect to the eavesdropper's channel approaches the output of the uniform distribution in variational distance. That is, each bin is a \emph{resolvability code} in the sense of Han and Verd\'u \cite{Han_Verdu_93}. In \cite{Bloch11, LuzziBloch11} it was shown that for discrete memoryless channels, resolvability-based random wiretap codes achieve strong secrecy; we have followed a similar approach for the Gaussian channel.

In the case when the target output distribution is capacity-achieving, a necessary condition for the bins to be resolvability codes is that the bin rate should be greater than the eavesdropper's channel capacity. Note that this is consistent with the condition (\ref{eq:binrate}): if $\Lambda_s$ is good for quantization, the entropy of the $\Lambda_s$-aliased noise
$\bar{\mathsf{W}}^n=[\mathsf{W}^n] \Mod \Lambda_s$ tends to the entropy of a white Gaussian noise with the same variance \cite{Zamir_Feder_96}, and $V(\Lambda_s)
\approx (2\pi e P)^{\frac{n}{2}}$, so the capacity $C(\Lambda_s,\sigma_e^2)$ of the eavesdropper's channel given by equation (\ref{mod_Lambda_capacity}) tends to $\frac{1}{2}\log 2\pi e P-\frac{1}{2} \log 2 \pi e
\sigma_e^2=\frac{1}{2}\log\SNR_e$. 
\end{rem}

\begin{rem}[Relation to Poltyrev's setting of infinite constellations]
Poltyrev initiated the study of infinite constellations in the
presence of Gaussian noise \cite{Poltyrev94}. In this setting,
although the standard channel capacity is meaningless (so he defined
generalized capacity), the secrecy capacity is finite. This is
because the secrecy capacity of the Gaussian wiretap channel as
$P\rightarrow \infty$ converges to a finite rate
$\frac{1}{2}\log(\frac{\sigma^2_e}{\sigma^2_b})$. Lattice codes can
not be better than this, so it is an upper bound. Even though we
considered a mod-$\Lambda_s$ channel in this section, we may enlarge
$\mathcal{V}(\Lambda_s)$ (i.e., increase $R'$ while fixing $R$) to
approach an infinite constellation. Since the upper bound (\ref{eq:leakage-bound-mod-Lambda}) on the mutual information of our proposed
scheme is independent of $V(\Lambda_s)$, the limit exists
as $V(\Lambda_s)\rightarrow \infty$. This corresponds to the case of
infinite constellations. Further, the achieved secrecy rate is only
a half nat away from the upper bound.
\end{rem}

\section{Gaussian Wiretap Channel}

Although the mod-$\Lambda$ channel has led to considerable insights,
there is no reason in real-world applications why the eavesdropper
would be restricted to use the modulo operation in the front end of her receiver. In this
section, we remove this restriction and solve the problem of the
Gaussian wiretap channel using lattice Gaussian coding.

\subsection{Channel Model} \label{AWGN_channel_model}

\begin{center}
\begin{figure}[hbt]

\begin{footnotesize}
\begin{tikzpicture}[
nodetype1/.style={
    rectangle,
    rounded corners,
    minimum width=26mm,
    minimum height=7mm,
    dashed,
    draw=black
},
nodetype2/.style={
    rectangle,
    rounded corners,
    minimum width=12mm,
    minimum height=7mm,
    draw=black
},
tip2/.style={-latex,shorten >=0.4mm}
]
\matrix[row sep=1.2cm, column sep=0.8cm, ampersand replacement=\&]{
\node (Alice) {\textsc{Alice}};  \& \node (encoder) [nodetype2]   {\textsc{encoder}}; \&
\node (W) {$\bigoplus$}; \&
\node (decoder) [nodetype2] {\textsc{decoder}}; \&
\node (Bob) {\textsc{Bob}};\\
\& (invisible) \&
\node (We)  {$\bigoplus$}; \&
\node (Eve) {\textsc{Eve}}; \&  \\};
\draw[->] (Alice) edge[tip2] node [above] {$\M,\mathsf{S}$} (encoder) ;
\draw[->] (encoder) edge[tip2] node [above] (X) {$\X^n$} (W) ;
\draw[->] (W) edge[tip2] node [above] {$\Y^n$}  (decoder) ;
\draw[->] (decoder) edge[tip2] node [above] {$\hat{\M}$} (Bob) ;
\draw[->] (We) edge[tip2] node [above] {$\Z^n$}  (Eve) ;
\draw[->,>=latex] (X) |- node [anchor=east] {} (We);
\node[above=0.5cm] (Nb) at (W.north) {$\mathsf{W}_b^n$};
\draw[->] (Nb) edge[tip2] (W);
\node[below=0.5cm] (Ne) at (We.south) {$\mathsf{W}_e^n$};
\draw[->] (Ne) edge[tip2] (We);
\end{tikzpicture}
\end{footnotesize}
\caption{The Gaussian wiretap channel.}
\label{fig:wiretapchannel}
\end{figure}
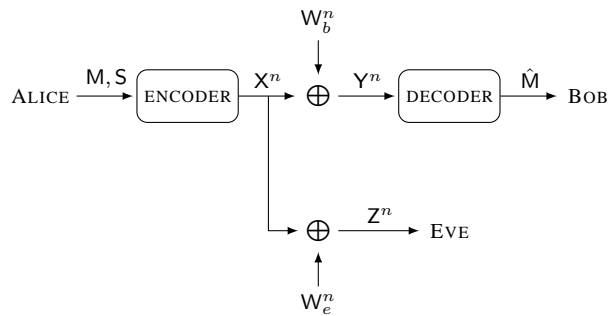
\end{center}

Let $\Lambda_e \subset \Lambda_b$ be $n$-dimensional lattices in $\R^n$ such that
\[
\frac{1}{n}\log \abs{\Lambda_b/\Lambda_e}=R.
\]
We consider the Gaussian wiretap channel depicted in Fig. \ref{fig:wiretapchannel}, whose outputs $\Y^n$ and $\Z^n$ at Bob and
Eve's end respectively are given by
\begin{equation} \label{Gaussian_wiretap}
\begin{cases}
\Y^n=\X^n + \mathsf{W}_b^n,\\
\Z^n=\X^n + \mathsf{W}_e^n,
\end{cases}
\end{equation}
where $\mathsf{W}_b^n$, $\mathsf{W}_e^n$ are $n$-dimensional Gaussian vectors with zero mean and variance $\sigma_b^2$, $\sigma_e^2$ respectively.
The transmitted codebook $\mathcal{C}$ must satisfy the average power constraint (\ref{power_constraint}). We denote this wiretap channel by $W(\sigma_b,\sigma_e,P)$. Again, let $\SNR_b=P/\sigma_b^2$ and $\SNR_e=P/\sigma_e^2$.

\subsection{Lattice Gaussian Coding} \label{AWGN_coding_scheme}
Consider a message set $\mathcal{M}_n=\{1,\ldots,e^{nR}\}$, and a
one-to-one function
$\phi: \mathcal{M}_n \to \Lambda_b/\Lambda_e$
which associates each message $m \in \mathcal{M}_n$ to a coset $\widetilde{\bm{\lambda}}_m\in \Lambda_b/\Lambda_e$. Again, one could choose the coset representative $\bm{\lambda}_m \in \Lambda_b \cap \mathcal{R}(\Lambda_e)$ for any fundamental region $\mathcal{R}(\Lambda_e)$. This is because the signal powers corresponding to different cosets will be nearly the same, as shown in the following. Good choices of fundamental region $\mathcal{R}(\Lambda_e)$ (e.g., the fundamental parallelepiped) can result in low-complexity implementation of the encoder and decoder, while choosing $\mathcal{V}(\Lambda_e)$ would require nearest-neighbor search \cite{Forney1}.
Note again that we make no \textit{a priori} assumption on the
distribution of~$m$.

In order to encode the message $m\in \mathcal{M}_n$, Alice
samples $\X_m^n$ from~$D_{\Lambda_e+\bm{\lambda}_m,\sigma_s}$ (as defined in Section~\ref{lattice_gaussian}); equivalently, Alice transmits $\bm{\lambda}+\bm{\lambda}_m$ where $\bm{\lambda}  \sim D_{\Lambda_e,\sigma_s, -\bm{\lambda}_m}$. The choice of the signal variance~$\sigma_s^2$ will be discussed later in this Section.

\begin{figure}[t]

\centering\centerline{\epsfig{figure=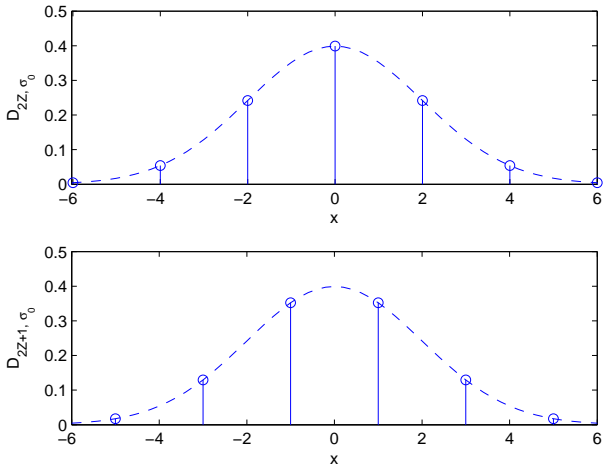,width=9cm}}

\caption{Lattice Gaussian coding (circle) over $2\mathbb{Z}$ and its coset $2\mathbb{Z}+1$ for $\sigma_s=2$. The profile (dashed) is the underlying continuous Gaussian distribution.}

\vspace{-0.5cm}

\label{fig:Gaussian-2Z}
\end{figure}

It is worth mentioning that the
distribution $D_{\Lambda_e+\bm{\lambda}_m,\sigma_s}$ is always centered at $\mathbf{0}$
for all bins. Fig. \ref{fig:Gaussian-2Z} illustrates the proposed lattice Gaussian coding using an example $\Lambda_e= 2\mathbb{Z}$ for $\sigma_s=2$. It is clear that both $D_{2\mathbb{Z},\sigma_s}$ and $D_{2\mathbb{Z}+1,\sigma_s}$ are centered at 0, sharing the same continuous Gaussian profile. This is key for the conditional output distributions corresponding to different $m$ to converge to the same distribution.


Lemma \ref{Micciancio_Regev_Lemma43} implies that if~$\epsilon_{\Lambda_e}\left(\sigma_s/\sqrt{\frac{\pi}{\pi-1/e}}\right) < 1/2$, then \[
\left| \mathbb{E}\left[ \norm{\X_m^n}^2\right]
- n\sigma_s^2 \right|
\leq \frac{2 \pi \epsilon_{\Lambda_e}\left(\sigma_s/\sqrt{\frac{\pi}{\pi-1/e}}\right)}{1-\epsilon_{\Lambda_e}\left(\sigma_s/\sqrt{\frac{\pi}{\pi-1/e}}\right)} \sigma_s^2,
\]
which is independent of $m$. Note that the overall input
distribution is a mixture of the densities of $\X_m^n$:
\begin{equation} \label{AWGN_input_distribution}
p_{\X^n}(\mathbf{x})=\sum_{m=1}^{e^{nR}} p_{\mathsf{M}}(m) p_{\X_m^n}(\mathbf{x}).
\end{equation}
Since the second moment in zero of a mixture of densities is the weighted sum of the second moments in zero of the individual densities, we have
\begin{equation} \label{AWGN_power}
\left|\frac{1}{n} \mathbb{E}\left[\norm{\X^n}^2\right]  - \sigma_s^2 \right| \leq \frac{2 \pi \epsilon_{\Lambda_e}\left(\sigma_s/\sqrt{\frac{\pi}{\pi-1/e}}\right)}{n\left[1-\epsilon_{\Lambda_e}\left(\sigma_s/\sqrt{\frac{\pi}{\pi-1/e}}\right)\right]}
\sigma_s^2.\end{equation}
We choose $\sigma_s^2 = P$ in order to satisfy the average power constraint
(\ref{power_constraint}) asymptotically (as $\epsilon_{\Lambda_e}\left(\sigma_s/\sqrt{\frac{\pi}{\pi-1/e}}\right) \to 0$). For convenience, let $\rho_b = \sigma_s^2/\sigma_b^2$ and $\rho_e = \sigma_s^2/\sigma_e^2$. It holds that $\rho_b \to \SNR_b$ and $\rho_e \to \SNR_e$ if $\epsilon_{\Lambda_e}\left(\sigma_s/\sqrt{\frac{\pi}{\pi-1/e}}\right) \to 0$.

\subsection{Achieving Strong Secrecy} \label{AWGN_secrecy}

We will now show that under suitable hypotheses, the conditional
output distributions at Eve's end converge in variational distance
to the same continuous Gaussian distribution, thereby achieving
strong secrecy.

Recall that Eve's channel transition probability is given by
\[
p_{\Z^n|\X^n}(\mathbf{z}|\bm{\lambda}_m+\bm{\lambda})=f_{\sigma_e, \bm{\lambda}_m+\bm{\lambda}}({\bf z}).
\]
Let $\tilde{\sigma}_e=\frac{\sigma_s\sigma_e}{\sqrt{\sigma_s^2+\sigma_e^2}}$. Lemma~\ref{Regev_Claim39} implies that
if~$\epsilon_{\Lambda_e}\left(\tilde{\sigma}_e\right)< \frac{1}{2}$, then:
\[
\V\left(p_{\Z^n|\M}(\cdot|m),f_{\sqrt{\sigma_s^2+\sigma_e^2}}\right)\leq 4
\epsilon_{\Lambda_e}\left(\tilde{\sigma}_e\right).
\]

An upper bound on the amount of leaked information then follows directly from Lemma~\ref{lemma2}.

\begin{ther}\label{theorem_leakage}
Suppose that the wiretap coding scheme described above is employed on the Gaussian wiretap channel~(\ref{Gaussian_wiretap}), and let $\varepsilon_n=\epsilon_{\Lambda_e}\left(\tilde{\sigma}_e\right)$.
Assume that~$\varepsilon_n<\frac{1}{2}$ for all~$n$.
Then the mutual information between the confidential message and the eavesdropper's signal is bounded as follows:
\begin{equation}
\I(\M;\Z^n) \leq 8\varepsilon_n nR
-8\varepsilon_n\log8\varepsilon_n
\end{equation}
\end{ther}

From Theorem~\ref{theorem_leakage}, we obtain a sufficient condition for a
sequence of nested lattice wiretap codes to achieve strong secrecy:

\begin{corol}
For any sequence of lattices
$\Lambda_e^{(n)}$ such that~$\epsilon_{\Lambda_e^{(n)}}\left(\tilde{\sigma}_e\right)=o\left(\frac{1}{n}\right)$
as $n \to \infty$, we have~$\I(\M,\Z^n) \to 0$.
\end{corol}

Note that $\tilde{\sigma}_e$ is smaller than both~$\sigma_e$ and~$\sigma_s$. The first inequality $\tilde{\sigma}_e < \sigma_e$ means that
\begin{itemize}
  \item Because of the monotonicity of the flatness factor (Remark \ref{monotonicity}), achieving strong secrecy on the Gaussian wiretap channel is a bit
more demanding than that on the mod-$\Lambda$ channel;
  \item Yet they are equally demanding at high SNR, since
  $\tilde{\sigma}_e\rightarrow
  \sigma_e$ as $\sigma_s\rightarrow \infty$.
\end{itemize}

The second inequality $\tilde{\sigma}_e < \sigma_s$ requires that $\epsilon_{\Lambda_e}(\sqrt{P})$
be small, which means that a minimum power $P$ is needed (specifically, $\sqrt{P}$ should be larger
than the smoothing parameter of~$\Lambda_e$).

\begin{rem}
Note that, similarly to the mod-$\Lambda$ case (Remark~\ref{resolvability_remark}) each bin of our strong secrecy scheme may be viewed as a resolvability code, and thus the bin rate must necessarily be above Eve's channel capacity. Indeed, the bin rate can be chosen to be quite close to this optimal value: note that for $\varepsilon_n$ in Theorem \ref{theorem_leakage} to vanish,
it suffices that
\begin{equation}\label{eq:strong-secrecy-condition}
\gamma_{\Lambda_e}\left(\tilde{\sigma}_e\right)=\frac{V(\Lambda_e)^{2/n}}{\tilde{\sigma}_e^2 }<2\pi
\end{equation}
for the mod-$p$ lattices of the first part of Theorem~\ref{theorem2}.
By Lemma~\ref{proposition4}, when $\varepsilon \triangleq \epsilon_{\Lambda_e}\left(\sigma_s/\sqrt{\frac{\pi}{\pi-1/e}}\right)<1$, the entropy rate of each bin satisfies
\begin{align*}
R' &\geq \log (\sqrt{2 \pi e}\sigma_s) - \frac{1}{n}\log {V(\Lambda_e)} - \frac{\varepsilon'}{n} \\
&> \log (\sqrt{2 \pi e}\sigma_s) - \frac{1}{2}\log \left(2\pi \frac{\sigma_s^2\sigma_e^2}{{\sigma_s^2+\sigma_e^2}}\right) - \frac{\varepsilon'}{n}\\
&= \frac{1}{2}\log \left(\frac{\sigma_s^2+\sigma_e^2}{{\sigma_e^2}}\right) + \frac{1}{2} - \frac{\varepsilon'}{n}\\
&= \frac{1}{2}\log \left(1+\rho_e\right) + \frac{1}{2} - \frac{\varepsilon'}{n}.
\end{align*}
where $\varepsilon'$ is defined in Lemma~\ref{proposition4}. Since $P\to \sigma_s^2$ as $\varepsilon \to 0$ (by~(\ref{AWGN_power})), we have $\rho_e \to \SNR_e$. Also, $\varepsilon' \to 0$ as $\varepsilon \to 0$. To make $\varepsilon \to 0$,
we only need an extra sufficient condition $\gamma_{\Lambda_e}\left({\sigma}_0/\sqrt{\frac{\pi}{\pi-1/e}}\right) < 2\pi$ for the mod-$p$ lattices of Theorem~\ref{theorem2}.
\end{rem}

\subsection{Achieving Reliability}

Now we show Bob can reliably decode the confidential message by using MMSE lattice decoding.
Consider the decoding scheme for Bob where he first decodes to the
fine lattice $\Lambda_b$, then applies the mod-$\Lambda_e$ operation
to recover the confidential message. We note that the distribution of Alice's signal can be approximated by $D_{\Lambda_b,\sigma_s}$, when the
confidential message is
uniformly distributed. More precisely, since Alice
transmits $\mathbf{x}\sim D_{\Lambda_e+\bm{\lambda}_m,\sigma_s}$, by Lemma \ref{GVP_Lemma27}, the density
$p_{\X^n}$ of $\mathbf{x}$ is close to the discrete
Gaussian distribution over~$\Lambda_b$, if $\bm{\lambda}_m\in \Lambda_b/\Lambda_e$ is
uniformly distributed. In fact, we have
$\V(p_{\X^n},D_{\Lambda_b,\sigma_s}) \leq
\frac{2\varepsilon}{1-\varepsilon}$ when~$\varepsilon \triangleq \epsilon_{\Lambda_e}(\sigma_s) <\frac{1}{2}$.

We will derive the maximum-a-posteriori
(MAP) decoding rule for decoding to $\Lambda_b$, assuming a discrete Gaussian distribution $D_{\Lambda_b,\sigma_s}$ over $\Lambda_b$. Since the lattice
points are not equally probable a priori in the lattice Gaussian
coding, MAP decoding is not the same as standard maximum-likelihood (ML)
decoding.

\begin{prop}[Equivalence between MAP decoding and MMSE lattice decoding]\label{equivalence}
Let $\mathbf{x} \sim D_{\Lambda_b,\sigma_s}$ be the input signaling
of an AWGN channel where the noise variance is $\sigma_b^2$. Then
MAP decoding is equivalent to Euclidean lattice decoding of~$\Lambda_b$ using a renormalized metric that is asymptotically close
to the MMSE metric.
\end{prop}

\begin{IEEEproof}
Bob receives $\mathbf{y}=\mathbf{x}+\mathbf{w}_b$. Thus the MAP
decoding metric is given by
\begin{align*}
\mathbb{P}(\mathbf{x}|\mathbf{y}) &=
\frac{\mathbb{P}(\mathbf{x},\mathbf{y})}{\mathbb{P}(\mathbf{y})}
\propto
\mathbb{P}(\mathbf{y}|\mathbf{x})\mathbb{P}(\mathbf{x}) \\
&\propto
\exp\left(-\frac{\norm{\mathbf{y}-\mathbf{x}}^2}{2\sigma_b^2}-\frac{\norm{\mathbf{x}}^2}{2\sigma_s^2}
\right)\\
&\propto
\exp\left(-\frac{1}{2}\left(\frac{\sigma_s^2+\sigma_b^2}{\sigma_s^2\sigma_b^2}
\norm{\frac{\sigma_s^2}{\sigma_s^2+\sigma_b^2}\mathbf{y}-\mathbf{x}}^2
\right)\right).
\end{align*}
Therefore,
\begin{align}
\arg\max_{\mathbf{x}\in\Lambda_b} \mathbb{P}(\mathbf{x}|\mathbf{y})
&= \arg\min_{\mathbf{x}\in\Lambda_b}
\norm{\frac{\sigma_s^2}{\sigma_s^2+\sigma_b^2}\mathbf{y}-\mathbf{x}}^2 \nonumber\\
&= \arg\min_{\mathbf{x}\in\Lambda_b}
\norm{\alpha{\mathbf{y}}-\mathbf{x}}^2 \label{eq:renormalized-metric}
\end{align}
where $\alpha=\frac{\sigma_s^2}{\sigma_s^2+\sigma_b^2}$ is known, thanks to (\ref{AWGN_power}), to be asymptotically close to the MMSE coefficient $\frac{P}{P+\sigma_b^2}$.
\end{IEEEproof}

Next we prove Bob's reliability for any secrecy rate close to the secrecy capacity. We use the $\alpha$-renormalized decoding metric (\ref{eq:renormalized-metric}), even if the confidential message is not necessarily uniformly distributed. In fact, the following proofs hold for any fixed message index~$m$.
Also note that no dither is required to achieve reliability. Indeed, as we will see, Regev's regularity lemma (Lemma~\ref{Regev_Claim39}) makes the dither unnecessary. This is because the equivalent noise will be asymptotically Gaussian.

Suppose Alice transmits message $m$, and Bob receives $\mathbf{y}=\mathbf{x}+\mathbf{w}_b=\bm{\lambda}+\bm{\lambda}_m+\mathbf{w}_b$ (with~$\bm{\lambda} \sim D_{\Lambda_e,\sigma_s, -\bm{\lambda}_m}$). From Proposition \ref{equivalence}, Bob
computes
\[
\hat{\bm{\lambda}}_m=\left[Q_{\Lambda_b}\left(\alpha{\mathbf{y}}\right)\right]\Mod \mathcal{R}(\Lambda_e).
\]
It is worth mentioning that since $Q_{\Lambda_b}\left(\alpha{\mathbf{y}}\right) \in \Lambda_b$, the $\Mod \mathcal{R}(\Lambda_e)$ operation is the remapping to cosets in $\Lambda_b/\Lambda_e$, which can be implemented easily \cite{Forney1}.

Recall the following properties of the $\Mod$ and
quantization operations. 
For all $\mathbf{a}, \mathbf{b} \in \R^n$, we have
\begin{align}
& [[\mathbf{a}] \Mod \mathcal{R}(\Lambda_e) +\mathbf{b}]\Mod \mathcal{R}(\Lambda_e)=[\mathbf{a}+\mathbf{b}] \Mod \mathcal{R}(\Lambda_e) \label{eq_34_NG}\\
& [Q_{\Lambda_b}(\mathbf{a})] \Mod \mathcal{R}(\Lambda_e)=\left[Q_{\Lambda_b}\left([\mathbf{a}] \Mod \mathcal{R}(\Lambda_e)\right)\right] \Mod \mathcal{R}(\Lambda_e). \label{eq_35_NG}
\end{align}
Using these properties, the output of Bob's decoder can be rewritten as
{
\allowdisplaybreaks
\begin{align*}
\hat{\bm{\lambda}}_m&=\left[Q_{\Lambda_b}\left(\mathbf{x}+\left(\alpha-1\right)\mathbf{x}+\alpha{\mathbf{w}_b}\right)\right]\Mod \mathcal{R}(\Lambda_e)\\
&=\left[Q_{\Lambda_b}\left(\left[\mathbf{x}+\left(\alpha-1\right)\mathbf{x}+\alpha{\mathbf{w}_b}\right]
\Mod \mathcal{R}(\Lambda_e)\right)\right]\Mod \mathcal{R}(\Lambda_e).
\end{align*}
}
Observe that since $\bm{\lambda} \in \Lambda_e$, we have
\begin{align*}
&\left[\mathbf{x}+\left(\alpha-1\right)\mathbf{x}+\alpha\mathbf{w}_b\right]
 \Mod \mathcal{R}(\Lambda_e) \\
=&\left[\bm{\lambda}_m+\left(\alpha-1\right)\mathbf{x}+\alpha\mathbf{w}_b\right] \Mod \mathcal{R}(\Lambda_e) \\
 = &\left[\bm{\lambda}_m+\tilde{\mathbf{w}}_b(m)\right] \Mod \mathcal{R}(\Lambda_e)
 \end{align*}
where
we have defined the equivalent noise
\[
\tilde{\mathbf{w}}_b(m)
=\left(\alpha-1\right)\mathbf{x}+\alpha\mathbf{w}_b.
\]
Therefore
\[
\hat{\bm{\lambda}}_m=\left[Q_{\Lambda_b}\left(\bm{\lambda}_m+\tilde{\mathbf{w}}_b(m)\right)\right]\Mod \mathcal{R}(\Lambda_e).
\] 

Let $p_{\tilde{\mathsf{W}}_b^n(m)}$ be the density of the equivalent noise $\tilde{\mathbf{w}}_b(m)$. Since $\mathbf{x}\sim D_{\Lambda_e+\bm{\lambda}_m,\sigma_s}$ and ${\mathbf{w}_b}$ is Gaussian, Lemma~\ref{Regev_Claim39} implies that for \emph{any fixed}~$m$, and randomizing over $\bm{\lambda}$, $p_{\tilde{\mathsf{W}}_b^n(m)}$ is very close to a continuous Gaussian distribution.
More precisely, applying Lemma~\ref{Regev_Claim39} with standard deviations $(\alpha-1)\sigma_s$
and $\alpha \sigma_b$, and defining $\tilde{\sigma}_b=\sqrt{(\alpha-1)^2\sigma_s^2+\alpha^2 \sigma_b^2}=\frac{\sigma_s\sigma_b}{\sqrt{\sigma_s^2+\sigma_b^2}}$, we have
\begin{equation}\label{eq:Gauss-noise}
\abs{p_{\tilde{\mathsf{W}}_b^n(m)}(\mathbf{w})-f_{\tilde{\sigma}_b}(\mathbf{w})} \leq 4\varepsilon'' f_{\tilde{\sigma}_b}(\mathbf{w}) \quad \forall {\bf w} \in \R^n,
\end{equation}
assuming that (recall $\rho_b=\sigma_s^2/\sigma_b^2$)
\[
\varepsilon''\triangleq \epsilon_{(1-\alpha)\Lambda_e}\left(\frac{(1-\alpha)\sigma_s}{\sqrt{1+1/\rho_b}}\right) = \epsilon_{\Lambda_e}\left(\frac{\sigma_s}{\sqrt{1+1/\rho_b}}\right) < \frac{1}{2}.
\]
Thus, if $\varepsilon'' \rightarrow 0$,  the equivalent noise is essentially statistically independent from~$m$, in the sense that it is very close to the distribution~$f_{\tilde{\sigma}_b}(\mathbf{w})$ that does not involve~$m$ at all.

\begin{ther} \label{reliability_theo}
Suppose $\SNR_b>e$ and $\frac{1+\SNR_b}{1+\SNR_e}>e$.
Then if $\Lambda_b^{(n)}$ is a sequence of AWGN-good lattices, and $\Lambda_e^{(n)}$ is a sequence of secrecy-good lattices, any strong secrecy rate~$R$ satisfying
\begin{equation} \label{total_rate}
R<\frac{1}{2}\log\left(\min\left\{ \frac{1+\SNR_b}{1+\SNR_e}, \SNR_b \right\}\right)-\frac{1}{2}
\end{equation}
is achievable on the Gaussian wiretap channel $W(\sigma_b,\sigma_e,P)$ using
the discrete Gaussian coding and MMSE-renormalized
Euclidean lattice decoding.
\end{ther}

\begin{IEEEproof}
The decoding error probability~$P_e(m)$ corresponding to the message $m$ is bounded from above as
\begin{multline*}
P_e(m) \leq \mathbb{P}\left\{Q_{\Lambda_b}\left(\bm{\lambda}_m+\tilde{\mathbf{w}}_b(m)\right) \neq \bm{\lambda}_m\right\}\\
=\mathbb{P}\left\{\tilde{\mathbf{w}}_b(m) \notin \mathcal{V}(\Lambda_b)\right\}.
\end{multline*}
Since in particular
\[
p_{\tilde{\mathsf{W}}_b^n(m)}(\mathbf{w})<(1+4\varepsilon'') f_{\tilde{\sigma}_b}(\mathbf{w}) \quad \forall {\bf w} \in \R^n,
\]
we find that
\begin{align*}
\mathbb{P}\left\{\tilde{\mathbf{w}}_b(m) \notin \mathcal{V}(\Lambda_b)\right\} \leq (1+4\varepsilon'') \cdot \mathbb{P}\left\{\hat{\mathbf{w}}_b \notin \mathcal{V}(\Lambda_b)\right\}
\end{align*}
where $\hat{\mathbf{w}}_b$ is i.i.d.\ Gaussian with variance
$\tilde{\sigma}_b^2$. Note that while the equivalent noise $\tilde{\mathbf{w}}_b(m)$ in general depends on $m$, the resulting bound on the error probability is independent of $m$.

From AWGN-goodness of $\Lambda_b$ \cite{ErezZamir04}, it follows that the decoding error probability $P_e$ tends to 0 exponentially fast if $\varepsilon''$ is bounded by a constant and if
\begin{equation}\label{eq:reliability-Bob}
\gamma_{\Lambda_b}\left(\tilde{\sigma}_b\right)=\frac{V(\Lambda_b)^{2/n}}{\tilde{\sigma}_b^2 }>2\pi e.
\end{equation}

On the other hand, since $\Lambda_e$ is secrecy-good, Theorem~\ref{theorem_leakage}
implies that a sufficient condition for the mod-$p$ lattices of Theorem~\ref{theorem2} to achieve strong secrecy is
\begin{equation}\label{eq:strong-secrecy}
\gamma_{\Lambda_e}\left(\tilde{\sigma}_e\right)=\frac{V(\Lambda_e)^{2/n}}{\tilde{\sigma}_e^2 }<2\pi.
\end{equation}

Combining (\ref{eq:reliability-Bob}) and (\ref{eq:strong-secrecy}),
we have that strong secrecy rates~$R$ satisfying
\begin{equation}\label{rate-bound1}
R=\frac{1}{n}\log{\frac{V(\Lambda_e)}{V(\Lambda_b)}}<\frac{1}{2} \log \left(\frac{1+\rho_b}{1+\rho_e}\right)-\frac{1}{2}
\end{equation}
are achievable.

Two extra conditions on the flatness factors are required. First, to make $\rho_b \to \SNR_b$ and  $\rho_e \to \SNR_e$, it suffices that $\epsilon_{\Lambda_e} \left(\sigma_s/\sqrt{\frac{\pi}{\pi-1/e}}\right) \to 0$ (by~(\ref{AWGN_power})). This condition can be satisfied by mod-$p$ lattices if
\begin{equation}\label{eq:power-condition}\nonumber
\gamma_{\Lambda_e}\left(\frac{\sigma_s}{\sqrt{\frac{\pi}{\pi-1/e}}}\right)=\frac{V(\Lambda_e)^{2/n}}{\frac{\sigma_s^2}{{\frac{\pi}{\pi-1}}} }<2\pi,
\end{equation}
which together with (\ref{eq:reliability-Bob}) limits the secrecy rate to
\begin{equation}\label{rate-bound2}
R<\frac{1}{2} \log \left(\frac{1+\rho_b}{{\frac{\pi}{\pi-1/e}}}\right)-\frac{1}{2}.
\end{equation}
The second condition $\epsilon_{\Lambda_e}\left(\frac{\sigma_s}{\sqrt{1+1/\rho_b}}\right) \to 0$ for the equivalent noise to be asymptotically Gaussian (by~(\ref{eq:Gauss-noise})) can be satisfied by mod-$p$ lattices if
\begin{equation}\label{eq:Gauss-condition}\nonumber
\gamma_{\Lambda_e}\left(\frac{\sigma_s}{\sqrt{1+1/\rho_b}}\right)=\frac{V(\Lambda_e)^{2/n}}{\frac{\sigma_s^2}{{1+1/\rho_b}} }<2\pi,
\end{equation}
which together with (\ref{eq:reliability-Bob}) limits the secrecy rate to
\begin{equation}\label{rate-bound3}
R<\frac{1}{2} \log {\rho_b}-\frac{1}{2}.
\end{equation}

Now, combining (\ref{rate-bound1})-(\ref{rate-bound3}) and considering a positive secrecy rate, we obtain (\ref{total_rate}) when $\SNR_b>e$ and $\frac{1+\SNR_b}{1+\SNR_e} >{e}$. Note that condition (\ref{rate-bound2}) has been absorbed in (\ref{total_rate}). Therefore, the theorem is proven.
\end{IEEEproof}

\begin{rem}
When $\SNR_b\cdot \SNR_e>1$, the first term of (\ref{total_rate}) is smaller. This leads to
\begin{equation} \label{total_rate_bound}
R<\frac{1}{2}\log(1+\SNR_b)-\frac{1}{2}\log(1+\SNR_e)-\frac{1}{2}
\end{equation}
which is within a half nat from the secrecy capacity.
\end{rem}

\begin{rem}
It can be checked that, in our framework, conventional (non-renormalized) minimum-distance lattice
decoding can only achieve strong secrecy rate up to
$$R < \frac{1}{2}\log \left(\SNR_b\right) - \frac{1}{2}\log \left(1+\SNR_e\right) -\frac{1}{2}.$$
This is because it requires
\[
\gamma_{\Lambda_b}(\sigma_b)=\frac{V(\Lambda_b)^{2/n}}{{\sigma}_b^2 }>2\pi e
\]
rather than (\ref{eq:reliability-Bob}). Therefore, MAP decoding or MMSE estimation allows to gain a constant 1 within
the logarithm of the first term.
\end{rem}

\begin{rem}
The existence of good wiretap codes for the Gaussian channel follows from Proposition \ref{existence_prop}.
In fact, this case is less demanding than the mod-$\Lambda_s$ channel there since no shaping lattice is needed.
We only need a sequence of nested lattices $\Lambda_e^{(n)} \subset
\Lambda_b^{(n)}$ where $\Lambda_e^{(n)}$ is secrecy-good (with respect to $\tilde{\sigma}_e$ rather than ${\sigma}_e$) and $\Lambda_b^{(n)}$ is AWGN-good.
\end{rem}

\begin{algorithm}[t]
\caption{\quad Klein Sampling Algorithm} \label{alg:Klein}
{\bf{Input}:} A basis $\mathbf{B} = [\mathbf{b}_1, \ldots \mathbf{b}_n]$ of $\Lambda$, $\sigma_s$, $\mathbf{c}$\\
{\bf{Output}:} ${\bm \lambda} \in \Lambda$ of distribution close to $D_{\Lambda,\sigma_s,{\bf c}}$\\
\vspace{-12pt}
\begin{algorithmic}[1]

\STATE ${\bm \lambda}=\mathbf{0}$

\FOR{$i = n, \ldots, 1$}

\STATE $\sigma_i=\sigma_s/\|\hat{\mathbf{b}}_i\|$, $c'_i=\langle \mathbf{c},\hat{\mathbf{b}}_i\rangle/\|\hat{\mathbf{b}}_i\|^2$

\STATE Sample $z_i$ from $D_{\mathbb{Z},\sigma_i,c'_i}$

\STATE $\mathbf{c}=\mathbf{c}-z_i \mathbf{b}_i$, ${\bm \lambda}={\bm \lambda}+z_i \mathbf{b}_i$

\ENDFOR

\RETURN ${\bm \lambda}$
\end{algorithmic}
\end{algorithm}

To encode, Alice needs an efficient algorithm to sample lattice points from the distribution~$D_{\Lambda_e+\bm{\lambda}_m,\sigma_s}$ over the coset $\Lambda_e+\bm{\lambda}_m$.
Without loss of generality, we discuss sampling from $D_{\Lambda-\mathbf{c},\sigma_s} = D_{\Lambda,\sigma_s,{\bf c}} - {\bf c}$ for some center $\mathbf{c}$. Fortunately, such an efficient algorithm exists when $\sigma_s$ is sufficiently large. More precisely, it was proven in~\cite{Gentry08} that Klein's
algorithm~\cite{Klein} samples from a distribution very close to~$D_{\Lambda,\sigma_s,{\bf c}}$
when~$\sigma_s$ is a bit larger than the norm of the possessed basis of~$\Lambda_e$.
Klein's sampling
algorithm is equivalent to a randomized version of successive interference
cancelation (SIC), and can be implemented in polynomial
complexity. Algorithm \ref{alg:Klein} shows the pseudo-code of Klein sampling, where $\hat{\mathbf{b}}_i$ ($i=1,\ldots,n$) are the Gram-Schmidt vectors of matrix~${\bf B}$. Note that Klein's algorithm has also been used in lattice decoding~\cite{LiuLS11},
to improve the performance of SIC.  The following result, adapted from~\cite{Gentry08},
ensures that the output distribution is close to~$D_{\Lambda,\sigma_s,{\bf c}}$.

\begin{lem} \label{Klein}
Given a basis~${\bf B}$ of an $n$-dimensional
lattice~$\Lambda$ and its Gram-Schmidt vectors $\hat{\mathbf{b}}_i$ ($i=1,\ldots,n$). Let $\eta_{\varepsilon}(\mathbb{Z})$ be the smoothing parameter of $\mathbb{Z}$ for $\varepsilon \leq \frac{1}{2}$. If~$\sigma_s \geq \eta_{\varepsilon}(\mathbb{Z})\cdot  \max_i\|\hat{\mathbf{b}}_i\|$, then for any~$\bf{c}$, the output of Klein's algorithm has distribution~$D'$ satisfying
\begin{equation}\label{Klein-distance}
\left|\frac{D'_{\Lambda,\sigma_s,{\bf c}}(\bm{\lambda})}{D_{\Lambda,\sigma_s,{\bf c}}(\bm{\lambda})}-1\right|
\leq
\left({1+4\varepsilon}\right)^n-1, \quad \forall \bm{\lambda} \in \Lambda.
\end{equation}
\end{lem}

It follows from \cite[Lemma 3.1]{Gentry08} that $\eta_{\varepsilon}(\mathbb{Z})\leq \omega(\sqrt{\log n})$ for some negligible $\varepsilon$. Thus, the condition $\sigma_s \geq \omega(\sqrt{\log n})\cdot  \max_i\|\hat{\mathbf{b}}_i\|$ is sufficient to ensure that the distance (\ref{Klein-distance}) vanishes.
One restriction of Lemma~\ref{Klein} is that in order to not require a too large~$\sigma_s$,
we need to possess a short basis for~$\Lambda_e$. Such a short basis may be found by lattice reduction, e.g., the LLL reduction \cite{LiuLS11}.

Obviously, the $L^1$ distance or statistical distance is also bounded as (\ref{Klein-distance}). The statistical distance is a convenient tool to analyze
randomized algorithms. An important property is that applying a deterministic or random function
to two distributions does not increase the statistical distance. This implies an algorithm behaves
similarly if fed two nearby distributions. More precisely, if the output satisfies a property with
probability~$p$ when the algorithm uses a
distribution~$D$, then the property is still satisfied with probability~$%
\geq p - \V(D,D')$ if fed~$D'$ instead of~$D$ (see~\cite[Chap.\ 8]%
{BK:Micciancio02}).

\section{Discussion}

In this paper, we have studied semantic security over the Gaussian wiretap channel using lattice codes.
The flatness factor serves as a new lattice parameter to measure information leakage in this setting.
It can tell whether a particular lattice is good or not for secrecy coding, and consequently provides
a design criterion of wiretap lattice codes. Since the message in encoded by the cosets (not the particular coset leaders),
mapping and demapping of the message can be implemented with low complexity. Consequently,
Bob's decoding complexity is essentially due to that of decoding the AWGN-good lattice.
While we have proved the existence
of secrecy-good mod-$p$ lattices, the explicit construction of practical secrecy-good lattices warrants an investigation.
Further work along the line of secrecy gain \cite{OSB11} may provide some hints on secrecy-good lattices.

The half-nat gap to the secrecy capacity is intriguing. It would be interesting to find out what happens in between,
and to further explore the relation between various lattice parameters.

\section*{Acknowledgments}
The authors would like to thank Matthieu Bloch, Guillaume Hanrot and Ram Zamir for helpful discussions.


%
%


\appendices

\section{Proof of Csisz\'ar's Lemma for Continuous Channels}
\label{Csis}

\begin{IEEEproof}
Note that in spite of the ambiguous
notation, here $p_{\Z}$ and $p_{\Z|\M=m}$ are densities on $\R^n$,
while $p_{\M}$ and $p_{\M|\Z=\mathbf{z}}$ are probability mass
functions on $\mathcal{M}_n$. We have {
\begin{align*}
d_{\av} & = \sum_{m \in \mathcal{M}_n} p_{\M}(m) \int_{\R^n} \abs{p_{\Z|\M=m}(z)-p_{\Z}(z)} dz\\
&=\sum_{m\in\mathcal{M}_n} \int_{\R^n} \abs{p_{\M|\Z=\mathbf{z}}(m)p_{\Z}(\mathbf{z})-p_{\M}(m)p_{\Z}(\mathbf{z})} d\mathbf{z}\\
&=\int_{\R^n} \sum_{m\in\mathcal{M}_n} \abs{p_{\M|\Z=\mathbf{z}}(m)-p_{\M}(m)}p_{\Z}(\mathbf{z}) d\mathbf{z}\\
&=\int_{\R^n} \V(p_{\M},p_{\M|\Z=\mathbf{z}})
d\mu \\
& =\int_{\R^n}\V_{\M} (\mathbf{z}) d\mu,
\end{align*}}%
where $\V_{\M}(\mathbf{z})=\V(p_{\M},p_{\M|\Z=\mathbf{z}})$ and $d\mu=p_{\Z}(\mathbf{z}) d\mathbf{z}$ is the probability
measure associated to~$\Z$.

By using  Lemma~2.7
in~\cite{CsiszarKorner81}, we obtain
{
\allowdisplaybreaks
\[
\h(\M)-\h(\M|\Z=\mathbf{z})\leq
\V_{\M} (\mathbf{z}) \log \frac{|\mathcal{M}_n|}{\V_{\M}(\mathbf{z})}.
\]
}

Multiplying by $p_{\Z}(\mathbf{z})$ and taking the integral, we find
\begin{align*}
\I(\M;\Z) &= \h(\M)-\h(\M|\Z) \\
&\leq  \int_{\R^n} \V_{\M}(\mathbf{z})
\log \frac{|\mathcal{M}_n|}{\V_{\M} (\mathbf{z})} d\mu \\
&=\int_{\R^n} \V_{\M} (\mathbf{z}) \log |\mathcal{M}_n|
d\mu-\int_{\R^n} \V_{\M} (\mathbf{z}) \log \V_{\M}(\mathbf{z}) d\mu.
\end{align*}
{From} Jensen's inequality, using the fact
that the function $t \mapsto t\log t$ is convex, we have that
\begin{align*}
\int_{\R^n} \V_{\M}(\mathbf{z}) \log & \V_{\M}(\mathbf{z})  d\mu  \\
&\geq \left(\int_{\R^n}\V_{\M} (\mathbf{z})d\mu\right)\log\left(\int_{\R^n}\V_{\M}(\mathbf{z})d\mu\right)\\
&=d_{\av} \log
d_{\av}.
\end{align*}
This completes the proof.
\end{IEEEproof}

\section{Existence of good nested lattices: \\
Proof of Proposition \ref{existence_prop}} \label{existence_prop_proof}

Let~$\mathcal{C}$ denote the set of $\mathbb{F}_p$-linear $(n,k)$ codes, and let~${C}$ be chosen uniformly at random from $\mathcal{C}$. Consider the corresponding Construction-A random lattice
\[
\tilde{\Lambda}_s=\frac{1}{p}{C}+\mathbb{Z}^n.
\]
By definition of the effective radius, we have:
$$p^k=\frac{\Gamma\left(\frac{n}{2}+1\right)}{\pi^{\frac{n}{2}}r_{\eff}(\tilde{\Lambda}_s)^n}.$$
We know from \cite[Theorem~5]{Erez_Litsyn_Zamir} that with high probability, the lattice~$\tilde{\Lambda}_s$ is Covering, quantization and AWGN-good if the following properties are satisfied:
\begin{enumerate}
\item[(i)] $\exists \beta<\frac{1}{2}: \ k \leq \beta n$,
\item[(ii)] $\lim\limits_{n \to \infty} \frac{k}{\log^2 n}=\infty$,
\item[(iii)] $\forall n: \ r_{\min} < r_{\eff}(\tilde{\Lambda}_s) < 2 r_{\min}$, where \[
r_{\min}=\min\left\{\frac{1}{4},\frac{(r_{\eff}(\tilde{\Lambda}_s))^2}{32n\sigma_b^2 E_P\left(\frac{r_{\eff}(\tilde{\Lambda}_s)}{\sqrt{n}\sigma_b}\right)}\right\}.
\]
\end{enumerate}
In the previous formula, $E_P$ denotes the Poltyrev exponent
\begin{equation}
E_P(\mu)=\begin{cases} \frac{1}{2} \left[(\mu-1)-\log \mu\right] & 1 < \mu \leq 2\\
\frac{1}{2} \log \frac{e\mu}{4} & 2 \leq \mu \leq 4\\
\frac{\mu}{8} & \mu \geq 4
\end{cases}
\end{equation}
where $\mu=\frac{\gamma_{\Lambda_s}(\sigma_b)}{2\pi e}$.
Property (iii) implies that the fundamental volume is bounded by
\begin{equation} \label{eq:r_min}
\frac{\pi^{\frac{n}{2}} (r_{\min})^n}{\Gamma\left(\frac{n}{2}+1\right)}  < V(\tilde{\Lambda}_s)=\frac{1}{p^{k}} <  \frac{\pi^{\frac{n}{2}} (2r_{\min})^n}{\Gamma\left(\frac{n}{2}+1\right)},
\end{equation}
which tends to $0$ faster than exponentially, since Euler's Gamma function grows faster than any exponential.
Given $(n,k)$ with~$k$ satisfying~(i) and~(ii),
consider $\tilde{p}(n,k)$ prime satisfying the condition (\ref{eq:r_min}). (The existence of such a prime number has been proven in \cite{Erez_Litsyn_Zamir}.)
	
As explained in \cite{Erez_Litsyn_Zamir} (end of Section III), in order to use~$\tilde{\Lambda}_s$ for power-constrained shaping it is necessary to scale it differently: we consider $\Lambda_s=a p\tilde{\Lambda}_s=\mathbf{B}_s \mathbb{Z}^n$
scaled so that its second moment
satisfies $\sigma^2(\Lambda_s)=P$.

Since $\Lambda_s$ is quantization-good, its normalized second moment
satisfies $G(\Lambda_s)=\frac{\sigma^2(\Lambda_s)}{V(\Lambda_s)^{\frac{2}{n}}}=\frac{P}{V(\Lambda_s)^{\frac{2}{n}}}
\to \frac{1}{2\pi e}$ as $n \to \infty$ \cite{ErezZamir04}.
Therefore
\[
V(\Lambda_s)^{\frac{2}{n}}=
\frac{P}{G(\Lambda_s)}\to 2\pi Pe.
\]
For large $n$, we have
\begin{equation} \label{eq:V_Lambda_s}
V(\Lambda_s)=a^n p^{n-k} \approx (2\pi e P)^{\frac{n}{2}}.
\end{equation}
Since $p^k$ grows superexponentially, so does~$p^{n-k}$ and we thus
have $a \to 0$ and
$a p \to \infty$ as $n \to \infty$.
If we set~$a$ in such a way that $V(\Lambda_s)$ is constant for~$a \to 0$ and~$p \to \infty$ (but may depend on~$n$),
then for each~$n$ we have a Minkowski-Hlawka type bound on the average behaviour of the theta series~$\Theta_{\Lambda_s}(\tau)$ (see Lemma~\ref{theta_average}). Fix $\delta_n>0$. For all~$n$, there exists $\bar{p}(n,k,\delta_n,\tau)$ such that for every prime $p >\bar{p}(n,k,\delta_n,\tau)$ and the corresponding~$a$,
\begin{equation} \label{Loeliger_bound}
\mathbb{E}\left[ \Theta_{\Lambda_s}(\tau)\right] \leq 1 + \delta_n+ \frac{1}{V(\Lambda_s) \tau^{\frac{n}{2}}}.
\end{equation}
The following lemma, proven in Appendix \ref{Appendix_proofs}, gives a more precise bound on the rate of convergence of the theta series to the Minkowski-Hlawka bound and guarantees that this choice of $p$ is compatible with (\ref{eq:r_min}).
\begin{lem} \label{rate_of_convergence}
There exists a sequence $\delta_n \to 0$ such that for sufficiently large~$n$,
we have~$\tilde{p}(n,k) > \bar{p}(n,k,\delta_n,y)$.
\end{lem}

Having defined the shaping lattice, we proceed with a nested code construction inspired by Section VII in \cite{ErezZamir04}.
Let $C_b$ be chosen uniformly in the ensemble  $\mathcal{C}_b$ of random linear $(n,k_b)$ codes over~$\mathbb{F}_q$, and denote by $\mathbf{A}_b$ its generator matrix.
We know from \cite{Nazer11} that if $\frac{n}{q}\to 0$, then the lattice
\[
\Lambda_b=\mathbf{B}_s\left(\frac{1}{q}C_b+\mathbb{Z}^n\right).
\]
is AWGN-good with high probability.
Let $k_e<k_b$, and let~$\mathbf{A}_e$ be the matrix whose columns are the first $k_e$ columns of~$\mathbf{A}_b$. This matrix generates an $(n,k_e)$ linear code $C_e$ over $\mathbb{F}_q$; note that averaging over the possible choices for $C_b$, this construction results in $C_e$ being a uniformly chosen $(n,k_e,q)$ linear code. We can consider the corresponding Construction-A lattice
\[
\Lambda_e=\mathbf{B}_s\left(\frac{1}{q}C_e+\mathbb{Z}^n\right).
\]
Clearly, we have $\Lambda_s \subseteq \Lambda_e \subseteq \Lambda_b$. As remarked in \cite{Nazer11}, there are many choices for $q$ and $k_e,k_b$ which ensure the properties
\begin{equation} \label{eq:bin_rate}
\begin{aligned}
&R'_n=\frac{1}{n} \log \frac{V(\Lambda_s)}{V(\Lambda_e)}=\frac{k_e}{n}\log q \to R', \\
&R_n=\frac{1}{n} \log \frac{V(\Lambda_e)}{V(\Lambda_b)}=\frac{k_b}{n} \log q \to R.
\end{aligned}
\end{equation}
For example we can choose $q$ to be the closest prime to $n\log n$ and define $k_e=\floor{nR'(\log q)^{-1}}$, $k_b=\floor{n(R+R')(\log q)^{-1}}$.
Consider the expectation over over the sets $\mathcal{C}$ and $\mathcal{C}_e$ of~$(n,k,p)$ and~$(n,k_e,q)$ linear codes.
By Proposition~\ref{expression_epsilon}, we have:
\begin{align}
&\lim_{n \to \infty} \mathbb{E}_{\mathcal{C}, \mathcal{C}_e} \left[ \epsilon_{\Lambda_e}(\sigma)\right] \notag\\
&=\lim_{n\to \infty} \left(\frac{\gamma_{\Lambda_e}(\sigma)}{2\pi}\right)^{\frac{n}{2}} \mathbb{E}_{\mathcal{C}}\left[\mathbb{E}_{\mathcal{C}_e}\left[\Theta_{\Lambda_e}\left(\frac{1}{2\pi\sigma^2}\right)\right]\right]-1. \label{double_expectation}
\end{align}
Let $f(\mathbf{x})=e^{-\pi \tau \norm{\mathbf{x}}^2}$, $\bar{\mathbf{v}}=\mathbf{v} \Mod q$, and $C_e^*=C_e \setminus \{\bf 0\}$. We have
{
\allowdisplaybreaks
\begin{align*}
&\mathbb{E}_{\mathcal{C}_e}\left[\Theta_{\Lambda_e}(\tau)\right]\\
&=\frac{1}{\abs{\mathcal{C}_e}}\sum_{C_e \in \mathcal{C}_e} \left(\sum_{\substack{\mathbf{v} \in \mathbb{Z}^n\\ \bar{\mathbf{v}}=0}} f\left(\frac{\mathbf{B}_s \mathbf{v}}{q}\right)+ \sum_{\substack{\mathbf{v} \in \mathbb{Z}^n\\ \bar{\mathbf{v}} \in C_e^*}} f\left(\frac{\mathbf{B}_s \mathbf{v}}{q}\right)\right)\\
&=\sum_{\mathbf{v} \in q\mathbb{Z}^n} f\left(\frac{\mathbf{B}_s \mathbf{v}}{q}\right)+\frac{q^{k_e}-1}{q^n-1} \sum_{\mathbf{v} \in \mathbb{Z}^n : \bar{\mathbf{v}}\neq 0} f\left(\frac{\mathbf{B}_s \mathbf{v}}{q}\right)\\
&=\sum_{\mathbf{v} \in \mathbb{Z}^n} f\left(\mathbf{B}_s \mathbf{v}\right)+\frac{q^{k_e}-1}{q^n-1} \sum_{\mathbf{v} \in \mathbb{Z}^n \setminus q\mathbb{Z}^n} f\left(\mathbf{B}_s \mathbf{v}/q\right)\\
&=\left(1-\frac{q^{k_e}-1}{q^n-1}\right) \Theta_{\Lambda_s}(\tau)+\frac{q^{k_e}-1}{q^n-1}\Theta_{\Lambda_s}\left(\frac{\tau}{q^2}\right).
\end{align*}
}
In the last equation we have used the equality $\Theta_{a \Lambda}(\tau)=\Theta_{\Lambda}(a^2 \tau)$.

We can now rewrite (\ref{double_expectation}) as
\begin{align*}
\lim_{n \to \infty} \left(\frac{\gamma_{\Lambda_e}(\sigma)}{2\pi}\right)^{\frac{n}{2}} \left(\mathbb{E}_{\mathcal{C}}\left[ \Theta_{\Lambda_s}(\tau)+\frac{1}{q^{n-k_e}}\Theta_{\Lambda_s}\left(\frac{\tau}{q^2}\right)\right]\right)-1
\end{align*}
where $\tau=\frac{1}{2\pi\sigma^2}$. Using the property (\ref{Loeliger_bound}), this can be bounded by
{\small
\begin{align*}
&\lim_{n \to \infty} \left(\frac{\gamma_{\Lambda_e}(\sigma)}{2\pi}\right)^{\frac{n}{2}} \left(1+\frac{(2\pi\sigma^2)^{\frac{n}{2}}}{V(\Lambda_s)}+\delta_n\right)\\
&+\lim_{n \to \infty} \left(\frac{\gamma_{\Lambda_e}(\sigma)}{2\pi}\right)^{\frac{n}{2}}\left(\frac{1}{q^{n-k_e}} \left(1+\frac{(2\pi\sigma^2q^2)^{\frac{n}{2}}}{V(\Lambda_s)}+\delta_n\right)\right)-1 \\
&\leq \lim_{n \to \infty} \left(\frac{\gamma_{\Lambda_e}(\sigma)}{2\pi}\right)^{\frac{n}{2}}\left(1+\frac{1}{e^{nR'}\left(\frac{\gamma_{\Lambda_e}(\sigma)}{2\pi}\right)^{\frac{n}{2}}}+\delta_n+\frac{1}{\left(\frac{\gamma_{\Lambda_e}(\sigma)}{2\pi}\right)^{\frac{n}{2}}}\right)\\
&=\lim_{n \to \infty} \left(\frac{\gamma_{\Lambda_e}(\sigma)}{2\pi}\right)^{\frac{n}{2}}(1+\delta_n)
\end{align*}
}
recalling that $e^{nR_n'}=q^{k_e}$ (see (\ref{eq:bin_rate})). Therefore $\Lambda_e$ is secrecy-good.

Further, we can show the majority of such lattices are secrecy-good. Fix $0<c \leq \frac{1}{2}$ and let $\delta=\frac{\left(\frac{\gamma_{\Lambda_e}(\sigma)}{2\pi}\right)^{\frac{n}{2}}(1+\delta_n)}{c}$. Then using Markov's inequality we get
$$ \mathbb{P}\left\{{\epsilon_{\Lambda_e}(\sigma)}\geq \delta \right\} \leq \frac{\mathbb{E}\left[\epsilon_{\Lambda_e}(\sigma)\right]}{\delta} \leq c$$
Therefore if $\gamma_{\Lambda_e}(\sigma)<2\pi$, the sequence ${\Lambda_e}^{(n)}$ is secrecy-good with probability greater than $1-c \geq \frac{1}{2}$.

To conclude, for $n$ large enough there exists a set of measure going to 1 in the ensemble $\mathcal{C} \times \mathcal{C}_b$ such that $\Lambda_s$ is quantization and AWGN-good and $\Lambda_b$ is AWGN-good \cite{ErezZamir04}, and a set of measure greater than $1/2$ in the same ensemble such that $\Lambda_e$ is secrecy-good. The intersection of these sets being non-empty, the existence of a good sequence of nested lattices follows as stated.

\section{Proofs of technical lemmas} \label{Appendix_proofs}

\subsection{Proof of Lemma~\ref{theta_average}} \label{app:MinkowskiHlawka}

Let $f(\mathbf{v})=e^{-\pi \tau \norm{\mathbf{v}}^2}$ for $\mathbf{v}
\in \mathbb{R}^n$ and fixed $\tau \in \R^+$, and denote by $C'$ the set
of all nonzero codewords of $C$. Following~\cite{Loeliger}, we have
\begin{align}
\frac{1}{|\mathcal{C}|} &\sum_{C\in \mathcal{C}}\sum_{\mathbf{v} \in
a\Lambda_C} f(\mathbf{v}) \notag \\
&= \frac{1}{|\mathcal{C}|}\sum_{C\in
\mathcal{C}}\left[\sum_{\mathbf{v} \in
\mathbb{Z}^n:\overline{\mathbf{v}}=0} f(a \mathbf{v}) +
\sum_{\mathbf{v} \in \mathbb{Z}^n:\overline{\mathbf{v}}\in C'}
f(a \mathbf{v})
\right] \notag \\
&= \sum_{\mathbf{v} \in \mathbb{Z}^n:\overline{\mathbf{v}}=0}
f(a \mathbf{v}) + \frac{p^k-1}{p^n-1} \sum_{\mathbf{v} \in
\mathbb{Z}^n:\overline{\mathbf{v}} \neq 0}
f(a \mathbf{v}) \label{eq:balance}\\
&= \sum_{\mathbf{v} \in a p\mathbb{Z}^n}f(\mathbf{v})+
\frac{p^k-1}{p^n-1}\left(\sum_{\mathbf{v} \in
a\mathbb{Z}^n}f(\mathbf{v})-\sum_{\mathbf{v} \in a
p\mathbb{Z}^n}f(\mathbf{v})\right) \label{eq:balance2}
\end{align}
where (\ref{eq:balance}) is due to the balance of~$\mathcal{C}$. We have
\begin{equation}\label{eq:condition1}
\sum_{\mathbf{v} \in a p\mathbb{Z}^n}f(\mathbf{v})=\Theta_{a p \mathbb{Z}^n} (\tau) \to 1
\end{equation}
for any $\tau>0$, since $a p \rightarrow \infty$ under the
conditions given. Moreover,
\begin{equation} \label{eq:second_term}
\frac{p^k-1}{p^n-1}\sum_{\mathbf{v} \in
a\mathbb{Z}^n}f(\mathbf{v}) \to V^{-1}\int_{\R^n} f(\mathbf{v})
d\mathbf{v}
\end{equation}
as $a \to 0$, $p \to \infty$ and $a^n p^{n-k}=V$ is fixed.
To see this, consider any sequence $a_{\ell} \to 0$ and define
$f_{\ell}(\mathbf{v})=f\left(a_{\ell}\nint{\frac{\mathbf{v}}{a_{\ell}}}\right)$,
then use Lebesgue's dominated convergence theorem, the functions
$f_{\ell}$ being dominated by $g(\mathbf{v})$ which is equal to $1$ if
$\mathbf{v} \in \left[-\frac{1}{2},\frac{1}{2}\right]^n$ and equal
to $e^{-\pi \tau \sum_{i=1}^n\left(\abs{v_i}-\frac{1}{2}\right)^2}$
otherwise.  Thus, we have
\begin{eqnarray}
\frac{1}{|\mathcal{C}|} \sum_{C\in \mathcal{C}}\sum_{\mathbf{v} \in
a\Lambda_C} f(\mathbf{v}) \rightarrow 1 + V^{-1}
\int_{\mathbb{R}^n} f(\mathbf{v}) d\mathbf{v}. \label{eq:balance3}
\end{eqnarray}
Since $\int_{\mathbb{R}^n} f(\mathbf{v}) d\mathbf{v} =
\tau^{-n/2}$, we obtain (\ref{theta-average}).

\begin{rem}
Although we are primarily concerned with the theta series, the average behavior (\ref{eq:balance3}) is more general and may be of independent interest. In fact, (\ref{eq:balance3}) holds as long as the function $f(\cdot)$ satisfies conditions (\ref{eq:condition1}) and (\ref{eq:second_term}).
\end{rem}

\subsection{Proof of the second part of Lemma \ref{GVP_Lemma27}} \label{Appendix_coset_distribution}
Let $\varepsilon=\epsilon_{\Lambda'}(\sigma)$.
From Lemma~\ref{GVP_Lemma26}, we have that $\forall \widetilde{\bm{\lambda}} \in [\Lambda/\Lambda']$,
\[
\frac{f_{\sigma,\widetilde{\bm{\lambda}}+\mathbf{c}}(\Lambda')}{f_{\sigma}(\Lambda')}
\in \left[\frac{1-\varepsilon}{1+\varepsilon}, 1 \right].
\]
Therefore, for all $\widetilde{\bm{\lambda}} \in [\Lambda/\Lambda']$:
\[
\frac{\abs{\Lambda/\Lambda'} \cdot f_{\sigma,\widetilde{\bm{\lambda}}+\mathbf{c}}(\Lambda')}{S} \in
\left[ \frac{1-\varepsilon}{1+\varepsilon}, \frac{1+\varepsilon}{1-\varepsilon}
\right],
\]
where~$S=\sum_{\widetilde{\bm{\lambda}} \in [\Lambda/\Lambda']} f_{\sigma, \widetilde{\bm{\lambda}}+\mathbf{c}}(\Lambda') \in \left[ \frac{1-\varepsilon}{1+\varepsilon}, 1\right]|\Lambda/\Lambda'|f_{\sigma}(\Lambda')$. As a consequence, for all $\bm{\lambda}' \in \Lambda'$:
{\allowdisplaybreaks
\begin{align*}
|D_{\Lambda,\sigma,\mathbf{c}}(\widetilde{\bm{\lambda}}&+\bm{\lambda}')-p_{\mathsf{L}+\mathsf{L}'}(\widetilde{\bm{\lambda}}+\bm{\lambda}')|\\ &=f_{\sigma,\mathbf{c}}(\widetilde{\bm{\lambda}}+\bm{\lambda}') \abs{\frac{1}{S} -\frac{1}{\abs{\Lambda/\Lambda'} f_{\sigma,\widetilde{\bm{\lambda}}+\mathbf{c}}(\Lambda')}}  \\
& \leq \frac{f_{\sigma,\mathbf{c}}(\widetilde{\bm{\lambda}}+\bm{\lambda}')}{S}
\max\left( \abs{1-\frac{1+\varepsilon}{1-\varepsilon}}, \abs{1-
\frac{1-\varepsilon}{1+\varepsilon}} \right) \\
&= \frac{2\varepsilon}{1-\varepsilon} D_{\Lambda,\sigma,\mathbf{c}}(\widetilde{\bm{\lambda}}+\bm{\lambda}'). \tag*{\IEEEQED}
\end{align*}
}

\subsection{Proof of Lemma \ref{Micciancio_Regev_Lemma43}}\label{Appendix_Lemma43}

Let $\mathbf{x}\sim D_{\Lambda,\sigma,\mathbf{c}}$. For convenience, we consider the case $s:=\sqrt{2\pi}\sigma=1$. The general case follows by scaling the lattice by a factor $s$. From \cite[p.14]{Micciancio05}, each component $x_i$ satisfies
\begin{equation}\label{}
  \left| \Exp[(x_i-c_i)^2] - \frac{1}{2\pi}\right| \leq
  \frac{\sum_{\mathbf{y}\in \Lambda^*}{y_i^2\cdot \rho(\mathbf{y})}}{1-\rho(\Lambda^*\setminus \mathbf{0})}
\end{equation}
where $\rho(\mathbf{y})=e^{-\pi \|\mathbf{y}\|^2}$. A bound $y_i^2 \leq \|\mathbf{y}\|^2 \leq e^{\|\mathbf{y}\|^2}$ was subsequently applied for each $i$ in \cite{Micciancio05}.
Here, we tighten this bound as follows. Firstly, we note that the following overall bound holds (by linearity)
\begin{equation}\label{bound52}
  \left| \Exp[\|\mathbf{x}-\mathbf{c}\| ^2] - \frac{n}{2\pi}\right| \leq
  \frac{\sum_{\mathbf{y}\in \Lambda^*}{\|\mathbf{y}\|^2\cdot \rho(\mathbf{y})}}{1-\rho(\Lambda^*\setminus \mathbf{0})},
\end{equation}
hence avoiding the multiple $n$ on the right-hand side. Secondly, since $y\leq e^{y/e}$, the numerator in (\ref{bound52}) can be bounded as
\[
\begin{split}
\sum_{\mathbf{y}\in \Lambda^*}{\|\mathbf{y}\|^2\cdot \rho(\mathbf{y})} &\leq \sum_{\mathbf{y}\in \Lambda^*\setminus \mathbf{0}}{e^{\|\mathbf{y}\|^2/e}\cdot e^{-\pi\|\mathbf{y}\|^2}} \\
&= \sum_{\mathbf{y}\in \Lambda^*\setminus \mathbf{0}}{e^{-(\pi-1/e)\|\mathbf{y}\|^2}} \\
&= \epsilon_{\Lambda}\left(\sigma/\sqrt{\frac{\pi}{\pi-1/e}}\right)
\end{split}
\]
rather than $\epsilon_{\Lambda}\left(\sigma/2\right)$.
Then Lemma \ref{Micciancio_Regev_Lemma43} follows.

It is possible to further reduce $\sqrt{\frac{\pi}{\pi-1/e}}$. Introduce a parameter $0<t\leq 1/e$, and let $Y$ be the larger solution of the two solutions to equation $y = e^{ty}$. Then the numerator in (\ref{bound52}) can be bounded by
\begin{equation}\label{}
\begin{split}
    & \sum_{\mathbf{y}\in \Lambda^*, \|\mathbf{y}\|\leq Y}{\|\mathbf{y}\|^2\cdot e^{-\pi\|\mathbf{y}\|^2}} +
     \sum_{\mathbf{y}\in \Lambda^*, \|\mathbf{y}\|>Y}{e^{t\|\mathbf{y}\|^2}\cdot e^{-\pi\|\mathbf{y}\|^2}} \\
    &\leq Y^2 \sum_{\mathbf{y}\in \Lambda^*, \|\mathbf{y}\|\leq Y} e^{-\pi\|\mathbf{y}\|^2} +
     \sum_{\mathbf{y}\in \Lambda^*, \|\mathbf{y}\|>Y}{e^{-(\pi-t)\|\mathbf{y}\|^2}} \\
    &\leq Y^2 \epsilon_{\Lambda}\left(\sigma\right) +
     \epsilon_{\Lambda}\left(\sigma/\sqrt{\frac{\pi}{\pi-t}}\right) \\
     &\leq \left(t^{-4}+1\right)\epsilon_{\Lambda}\left(\sigma/\sqrt{\frac{\pi}{\pi-t}}\right) \nonumber
\end{split}
\end{equation}
where the last step is because $t=\log (Y) /Y\leq 1/\sqrt{Y}$. Thus, for a small but fixed value of $t$, the coefficient $\sqrt{\frac{\pi}{\pi-t}}$ can be very close to $1$, at the cost of a large constant $t^{-4}+1$.

\subsection{Proof of Lemma \ref{rate_of_convergence}}

We study more explicitly the rate of convergence, by going back to the expression (\ref{eq:balance2}) in the proof of  Lemma~\ref{theta_average}. We can rewrite it as
\begin{align*}
&\left(1-\frac{p^k-1}{p^n-1}\right) \left(\Theta_{\mathbb{Z}^n}(a^2p^2\tau)\right) +\frac{p^k-1}{p^n-1} \left(\Theta_{\mathbb{Z}^n}(a^2 \tau)\right)\\
&=\left(1-\frac{p^k-1}{p^n-1}\right) \left(\Theta_{\mathbb{Z}}(a^2p^2\tau)\right)^n +\frac{p^k-1}{p^n-1} \left(\Theta_{\mathbb{Z}}(a^2 \tau)\right)^n
\end{align*}
From the bound
\begin{align*}
\int_{\R} e^{-\tau z^2} dz&= 2 \int_0^\infty e^{-\tau z^2} dz \\
&\leq \Theta_{\mathbb{Z}}(\tau) = 1+2 \sum_{z \geq 1} e^{-yz^2} \\
& \leq 1 +2\int_0^\infty e^{-\tau z^2} dz=1+ \int_{\R} e^{-\tau z^2} dz,
\end{align*}
and recalling that $a^n p^{n-k}=V$, we find that
\begin{align*}
&\frac{1}{p^{n-k}} \left(\Theta_{\mathbb{Z}}(a^2 \tau)\right)^n \leq \frac{a^n}{V} \left(1+\frac{1}{a} \int_{\R} e^{-\tau z^2} dz\right)^n\\
&=\frac{1}{V} \int_{\R^n} e^{-\tau\norm{\mathbf{v}}^2} d\mathbf{v}+O\left(\frac{n}{V^{1-\frac{1}{n}}p^{1-\frac{k}{n}}}\right),
\end{align*}
while the lower bound is simply
$$\frac{1}{p^{n-k}} \left(\Theta_{\mathbb{Z}}(a^2 \tau)\right)^n \geq \frac{1}{V} \int_{\R^n} e^{-\tau\norm{\mathbf{v}}^2} d\mathbf{v}.$$
Similarly, we have
$$1 \leq \left(\Theta_{\mathbb{Z}}(a^2p^2\tau)\right)^n \leq 1 + \frac{1}{V^{\frac{1}{n}}p^{\frac{k}{n}}}n\int_{\R} e^{-\tau z^2} dz +o\left(\frac{n^{\frac{1}{n}}}{p^{\frac{k}{n}}}\right).$$
It is not hard to see that the sequence $\tilde{p}(n,k)$ defined by (\ref{eq:r_min}) ensures (more than exponentially fast) convergence. \quad \quad \quad $\IEEEQED$

\footnotesize
\bibliographystyle{IEEEtran}
\bibliography{IEEEabrv,lingbib}

%





\end{document}